\documentclass[iop]{emulateapj}
\pdfoutput=1
\usepackage{graphicx}
\usepackage{epstopdf}

\slugcomment{Accepted For Publication in The Astrophysical Journal: January 23, 2018}

\shorttitle{Combined Dipole, Quadrupole and Octupole Fields}
\shortauthors{A. Finley \& S. Matt}

\begin{document}


\title{The Effect of Combined Magnetic Geometries on Thermally Driven Winds II: \\
Dipolar, Quadrupolar and Octupolar Topologies}


\author{Adam J. Finley* \& Sean P. Matt}
\affil{University of Exeter, Department of Physics \& Astronomy,
              Devon, Exeter, EX4 4QL, UK}
\email{*af472@exeter.ac.uk}



\begin{abstract}
During the lifetime of sun-like or low mass stars a significant amount of angular momentum is removed through magnetised stellar winds. This process is often assumed to be governed by the dipolar component of the magnetic field. However, observed magnetic fields can host strong quadrupolar and/or octupolar components, which may influence the resulting spin-down torque on the star. In Paper I, we used the MHD code PLUTO to compute steady state solutions for stellar winds containing a mixture of dipole and quadrupole geometries. We showed the combined winds to be more complex than a simple sum of winds with these individual components. This work follows the same method as Paper I, including the octupole geometry which increases the field complexity but also, more fundamentally, looks for the first time at combining the same symmetry family of fields, with the field polarity of the dipole and octupole geometries reversing over the equator (unlike the symmetric quadrupole). We show, as in Paper I, that the lowest order component typically dominates the spin down torque. Specifically, the dipole component is the most significant in governing the spin down torque for mixed geometries and under most conditions for real stars. We present a general torque formulation that includes the effects of complex, mixed fields, which predicts the torque for all the simulations to within $20\%$ precision, and the majority to within $\approx5\%$. This can be used as an input for rotational evolution calculations in cases where the individual magnetic components are known.
\end{abstract}




\section{Introduction}
Cool stars are observed to host global magnetic fields which are embedded within their outer convection zones \citep[][]{reiners2012observations}. Stellar magnetism is driven by an internal dynamo which is controlled by the convection and stellar rotation rate, the exact physics of which is still not fully understood (see review by \citealp{brun2017magnetism}). As observed for the Sun, plasma escapes the stellar surface, interacting with this magnetic field and forming a magnetised stellar wind that permeates the environment surrounding the star \citep[][]{cranmer2017origins}.Young main sequence stars show a large spread in rotation rates for a given mass. As a given star ages on the main sequence, their stellar wind removes angular momentum, slowing the rotation of the star \citep{schatzman1962theory, weber1967angular, mestel1968magnetic}. This in turn reduces the strength of the magnetic dynamo process, feeding back into the strength of the applied stellar wind torque. This relationship leads to a convergence of the spin rates towards a tight mass-rotation relationship at late ages, as stars with faster rotation incur larger spin down torques and vice versa for slow rotators. This is observed to produce a simple relation between rotation period and stellar age \citep[$\Omega_*\propto t^{-0.5},$][]{skumanich1972time}, which is approximately followed, on average \citep{soderblom1983rotational} over long timescales.

With the growing number of observed rotation periods \citep{irwin2009ages, agueros2011factory, meibom2011color, mcquillan2013measuring, bouvier2014angular, stauffer2016rotation, 2017ApJ...835...16D}, an increased effort has been channelled into correctly modelling the spin down process \citep[e.g.][]{reiners2012radius, gallet2013improved, van2013fast, brown2014metastable, matt2015mass, gallet2015improved, amard2016rotating, blackman2016minimalist, see2017open}, as it is able to test our understanding of basic stellar physics and also date observed stellar populations. 

The process of generating stellar ages from rotation is referred to as Gyrochronology, whereby a cluster's age can be estimated from the distribution of observed rotation periods \citep{barnes2003rotational, meibom2009stellar, barnes2010simple, delorme2011stellar, van2013fast}. This requires an accurate prescription of the spin down torques experienced by stars due to their stellar wind, along with their internal structure and properties of the stellar dynamo. Based on results from magnetohydrodynamic (MHD) simulations, parametrised relations for the stellar wind torque are formulated using the stellar magnetic field strength, mass loss rate and basic stellar parameters (\citealp{mestel1984angular}; \citealp{kawaler1988angular}; \citealp{matt2008accretion}; \citealp{matt2012magnetic}; \citealp{ud2009dynamical}; \citealp{pinto2011coupling}; \citealp{reville2015effect}). The present work focusses on improving the modelled torque on these stars due to their magnetised stellar winds, by including the effects of combined magnetic geometries. 

Magnetic field detections from stars, other than the Sun, are reported over 30 years ago via Zeeman broadening observations \citep{robinson1980observations, marcy1984observations, gray1984measurements}, which has since been used on a multitude of stars \citep[e.g.][]{saar1990magnetic, johns2000measurements}. This technique, however, only allows for an average line of sight estimate of the unsigned magnetic flux and provides no information about the geometry of the stellar magnetic field (see review by \cite{reiners2012observations}). More recently, the use of Zeeman Doppler Imaging (ZDI), a tomographic technique capable of providing information about the photospheric magnetic field of a given star, enables the observed field to be broken down into individual spherical harmonic contributions \citep[e.g.][]{hussain2002coronal, donati2006large, donati2008magnetic, morin2008stable, morin2008large, petit2008toroidal, fares2009magnetic, morgenthaler2011direct, vidotto2014stellar, jeffers2014e, see2015energy, saikia2016solar,see2016connection, folsom2016evolution, hebrard2016modelling, see2016studying,kochukhov2017surface}. This allows the 3D magnetic geometry to be recovered, typically using a combination of field extrapolation and MHD modelling \citep[e.g.][]{vidotto2011understanding, cohen2011dynamics, garraffo2016space, reville2016age, alvarado2016simulating, nicholson2016temporal, do2016magnetic}. 

Pre-main sequence stars, observed with ZDI, show a variety of multipolar components, typically dependent on the internal structure of the host star \citep{gregory2012can, hussain2013role}. Many of these objects show an overall dipolar geometry with an accompanying octupole component \citep[e.g.][]{donati2007magnetic, gregory2012can}. The addition of dipole and octupole fields has been explored analytically, for these stars, and is shown to impact the disk truncation radius along with the topology and field strength of accretion funnels \citep{gregory2011analytic, gregory2016multipolar}. For main sequence stellar winds, the behaviour of combined magnetic geometries has yet to be systematically explored. Our closest star, the Sun, hosts a significant quadrupolar contribution during the solar activity cycle maximum which dominates the large scale magnetic field geometry along with a small dipole component \citep{derosa2012solar, brun2013rotation}. The impact of these mixed geometry fields on the spin down torque generated from magnetised stellar winds remains uncertain.

It is known that the magnetic field stored in the lowest order geometries, e.g. dipole, quadrupole \& octupole, has the slowest radial decay and therfore governs the strength of the magnetic field at the Alfv\'en surface (and thus it's size and shape). With the cylindrical extent of the Alfv\'en surface being directly related to the efficiency of the magnetic braking mechanism, it is this global field strength and geometry that is required to compute accurate braking torques in MHD simulations \citep{reville2015effect, reville2016age}. However, the effect of the higher order components on the acceleration of the wind close in to the star may not be non-negligible \citep{cranmer2005generation, cohen2009effect}. Additionally, the small scale surface features described by these higher order geometries (e.g. star spots and active regions) will play a vital role in modulating the chromospheric activity \citep[e.g.][]{testa2004density, aschwanden2006physics,gudel2007sun, garraffo2013effect}, which is often assumed to be decoupled from the open field regions producing the stellar wind. Models such as the AWESOM \citep{van2014alfven} include this energy dissipation in the lower corona, and are able to match observed solar parameters well. Work by \cite{pantolmos2017magnetic}, shows how this additional acceleration can be accounted for globally within their semi-analytic formulations.

Previous works have aimed to understand the impact of more complex magnetic geometries on the rotational evolution of sun-like stars. \cite{holzwarth2005impact} examined the effect of non-uniform flux distributions on the magnetic braking torque, investigating the latitudinal dependence of the stellar wind produced within their MHD simulations. Similarly, \cite{garraffo2016missing} included magnetic spots at differing latitudes and examined the resulting changes to mass loss rate and spin down torque. The effectiveness of the magnetic braking from a stellar wind is found to be reduced for higher order magnetic geometries \citep{garraffo2015dependence}. This is explained in \cite{reville2015effect} as a reduction to the average Alfv\'en radius, which acts mathematically as a lever arm for the applied braking torque. \cite{finley2017dipquad}, hereafter Paper I, continue this work by discussing the morphology and braking torque generated from combined dipolar and quadrupolar field geometries using ideal MHD simulations of thermally driven stellar winds. In this current work, we continue this mixed field investigation by including combinations with an octupole component.

Section 2 introduces the simulations and the numerical methods used, along with our parametrisation of the magnetic field geometries and derived simulation properties. Section 3 explores the resulting relationship of the average Alfv\'en radius with increasing magnetic field strength for pure fields, and generic combinations of axisymmetric dipole, quadrupole or octupole geometries. Section 4 uses the decay of the unsigned magnetic flux with distance to explain observed behaviours in our Alfv\'en radii relations, analysis of the open magnetic flux in our wind solutions follows with a singular relation for predicting the average Alfv\'en radius based on the open flux. Conclusions and thoughts for future work can be found in Section 5. 

\section{Simulation Method and Numerical Setup}
As in Paper I, we use the PLUTO MHD code \citep{mignone2007pluto, mignone2009pluto} with a spherical geometry to compute 2.5D (two dimensions, $r$, $\theta$, and three vector components, $r$, $\theta$, and $\phi$) steady state wind solutions for a range of magnetic geometries. 

The full set of ideal MHD equations are solved, including the energy equation and a closing equation of state. The internal energy density $\epsilon$ is given by $\rho\epsilon=p/(\gamma-1)$, where $\gamma$ is the ratio of specific heats. This general set of equations is capable of capturing non-adiabatic processes, such as shocks, however the solutions found for our steady-state winds generally do not contain these. For a gas comprised of protons and electrons $\gamma$ should take a value of 5/3, however we decrease this value to 1.05 in order to reproduce the observed near isothermal nature of the solar corona \citep{steinolfson1988density} and a terminal speed consistent with the solar wind. This is done, such that on large scales the wind follows the polytropic approximation, i.e. the wind pressure and density are related as, $p\propto \rho^{\gamma}$ \citep{parker1965dynamical, keppens1999numerical}.  The reduced value of $\gamma$ has the effect of artificially heating the wind as it expands, without an explicit heating term in our equations. 

We adopt the numerics used in Paper I, except that we modify the radial discretisation of the computational mesh. Instead of a geometrically stretched radial grid as before, we now employ a stepping ($dr$) that grows logarithmically. The domain extent remains unchanged, from one stellar radius ($R_*$) to 60$R_*$, containing $N_r\times N_{\theta}=256\times512$ grid cells. This modification produces a more consistent aspect ratio between $dr$ and $rd\theta$ over the whole domain, which marginally increases our numerical accuracy and stability. 

Characteristic speeds such as the surface escape speed and Keplerian speed, $v_{\text{esc}}$, $v_{\text{kep}}$, the equatorial rotation speed, $v_{\text{rot}}$, along with the surface adiabatic sound speed, $c_{\text{s}}$, and Alfv\'en speed, $v_{\text{A}}$, are given,
\begin{equation}
v_{\text{esc}}=\sqrt{\frac{2GM_*}{R_*}}=\sqrt{2}v_{\text{kep}},
\end{equation}
where, $G$ is the gravitational constant, $R_*$ is the stellar radius and $M_*$ is the stellar mass,
\begin{equation}
v_{\text{rot}}=\Omega_* R_*,
\end{equation}
where $\Omega_*$ is the angular stellar rotation rate (which is assumed to be in solid body rotation),
\begin{equation}
c_{\text{s}}=\sqrt{\frac{\gamma p_*}{\rho_*}},
\label{polytropic}
\end{equation}
where $\gamma$ is the polytropic index, $p_*$ and $\rho_*$ are the gas pressure and mass density at the stellar surface respectively,
\begin{equation}
v_{\text{A}}=\frac{B_*}{\sqrt{4\pi\rho_*}},
\end{equation}
where $B_*$ is the characteristic polar magnetic field strength (see Section 2.1). 

  \begin{table}
\caption{Fixed Simulation Parameters}
\label{Constants}
\center
\setlength{\tabcolsep}{1pt}
    \begin{tabular}{c|cc}
        \hline\hline
Parameter	&	Value	&	Description	\\	\hline
$\gamma$	&	1.05 &	Polytropic Index	\\
$c_{\text{s}}/v_{\text{esc}}$	&	0.25	&	Surface Sound Speed/ Escape Speed	\\
$f$	&	4.46E-03	&	Fraction of Break-up Rotation	\\ \hline
    \end{tabular}
\end{table}

We set an initial wind speed within the domain using a spherically symmetric Parker wind solution \citep{parker1965dynamical}, with the ratio of the surface sound speed to the escape speed $c_{\text{s}}/v_{\text{esc}}$ setting the base wind temperature in such a way as to represent a group of solutions for differing gravitational field strengths. The same normalisation is applied to the surface magnetic field strength with $v_{\text{A}}/v_{\text{esc}}$, and the surface rotation rate using $f=v_{\text{rot}}/v_{\text{kep}}$, such that each wind solution represents a family of solutions that  can be applied to a range of stellar masses. The same system of input parameters are used by many previous authors \citep[e.g.][]{matt2008accretion, matt2012magnetic, reville2015effect, pantolmos2017magnetic}. For this study we fix the wind temperature and stellar rotation at the values tabulated in Table \ref{Constants}. 

A background field corresponding to our chosen potential magnetic field configuration (see Section \ref{Magconfig}) is imposed over the initial wind solution and then all quantities are evolved to a steady state solution by the PLUTO code. The boundary conditions are enforced, as in Paper I, at the inner radial boundary (stellar surface) which are appropriate to give a self consistent wind solution for a rotating magnetised star. A fixed surface magnetic geometry is therefore maintained along with solid body rotation.

The use of a polytropic wind produces solutions which are far more isotropic than observed for the Sun \citep{vidotto2009three}. The velocity structure of the solar wind is known to be largely bimodal, having a slow and fast component which originate under different circumstances \citep{fisk1998slow, feldman2005sources, riley2006comparison}. This work and previous studies using a polytropic assumption aim to model the globally averaged wind which can be more generally applied to the variety of observed stellar masses and rotation periods. More complex wind driving and heating physics are needed in order to reproduce the observed velocity structure of the solar wind, however they are far harder to generalise for other stars \citep{cranmer2007self, pinto2016flux}.
 
\subsection{Magnetic Field Configurations}\label{Magconfig}
The magnetic geometries considered in this work include dipole, quadrupole and octupole combinations, with different field strengths and in some cases relative orientations. As in Paper I, we describe the mixing of different field geometries using the ratio of the polar field strength in a given component to the total field strength. Care is taken to parametrise the field combinations due to the behaviour of the two equatorially antisymmetric components, dipole and octupole, at the poles.

We generalise the ratio defined within Paper I for each component such that,
\begin{equation}
\mathcal{R}_x=\frac{B_*^{l=x}}{|B_*^{l=1}|+|B_*^{l=2}|+|B_*^{l=3}|}=\frac{B_*^{l=x}}{B_*},
\label{rvalue}
\end{equation}
where in this work, $l$ is the principle spherical harmonic number and $x$ can value 1, 2 or 3 for dipole, quadrupole or octupole fields. The polar field strength of a given component is written as $B_*^{l=x}$ and the $B_*=|B_*^{l=1}|+|B_*^{l=2}|+|B_*^{l=3}|$ is a characteristic field strength. The polar field strengths in the denominator are given with absolute values, because we are interested in the characteristic strength of the combined components, which are the same for aligned and anti-aligned fields. Therefore summing the absolute value of the ratios produces unity,
\begin{equation}
\sum_{l=1}^3|\mathcal{R}_l| = 1,
\end{equation}
which allows the individual values of $\mathcal{R}_{\text{dip}},\mathcal{R}_{\text{quad}}$ and $\mathcal{R}_{\text{oct}}$ ($\equiv \mathcal{R}_{1},\mathcal{R}_{2}$ and $\mathcal{R}_{3}$) to range from 1 to -1 (north pole positive or negative), with the absolute total remaining constant. We define the magnetic field components using these ratios and the Legendre polynomials $P_{lm}$, which for the axisymmetric ($m=0$) field components can be written,
\begin{eqnarray}
B_r(r,\theta)&=&B_*\sum_{l=1}^3\mathcal{R}_l P_{l0}(cos\theta)\bigg(\frac{R_*}{r}\bigg)^{l+2},\\
B_{\theta}(r,\theta)&=&B_*\sum_{l=1}^3\frac{1}{l+1}\mathcal{R}_l P_{l1}(cos\theta)\bigg(\frac{R_*}{r}\bigg)^{l+2}.
\end{eqnarray}
The northern polar magnetic field strengths for each components are given by,
\begin{equation}
B_*^{l=1}=\mathcal{R}_{\text{dip}}B_*,\;  B_*^{l=2}=\mathcal{R}_{\text{quad}}B_*,\;  B_*^{l=3}=\mathcal{R}_{\text{oct}}B_*,
\end{equation}

The relative orientation of the magnetic components is controlled throughout this work by setting the dipole and quadrupole fields ($B_*^{l=1}$ and $B_*^{l=2}$) to be positive at the northern stellar pole. The octupole component ($B_*^{l=3}$) is then combined with the dipolar and quadruplar components using either a positive or negative strength on the north pole, which we define as the aligned and anti-aligned cases respectively.

  \begin{table*}
\caption{Input Parameters and Results from Simulations with One \& Two Magnetic Components}
\label{Parameters}
\center
\setlength{\tabcolsep}{1pt}
    \begin{tabular}{ccccccc|ccccccc}
        \hline\hline
Case	&	$\mathcal{R}_{\text{dip}}|\mathcal{R}_{\text{quad}}|\mathcal{R}_{\text{oct}}$	&	$v_{\text{A}}/v_{\text{esc}}$	&	$\langle R_{\text{A}}\rangle/R_*$	&	$\Upsilon$	&	$\Upsilon_{\text{open}}$	&	$\langle v(R_{\text{A}})\rangle/v_{\text{esc}} $	&	Case	&	$\mathcal{R}_{\text{dip}}|\mathcal{R}_{\text{quad}}|\mathcal{R}_{\text{oct}}$	&	$v_{\text{A}}/v_{\text{esc}}$	&	$\langle R_{\text{A}}\rangle/R_*$	&	$\Upsilon$	&	$\Upsilon_{\text{open}}$	&	$\langle v(R_{\text{A}})\rangle/v_{\text{esc}} $	\\	\hline
1	&	$1.0|0.0|0.0$	&	0.5	&	5.0	&	185	&	1460	&	0.22	&	65	&	$0.5|0.0|0.5$	&	0.5	&	3.8	&	203	&	648	&	0.17	\\	
2	&	$1.0|0.0|0.0$	&	1.0	&	6.9	&	735	&	3540	&	0.29	&	66	&	$0.5|0.0|0.5$	&	1.0	&	4.9	&	705	&	1380	&	0.22	\\	
3	&	$1.0|0.0|0.0$	&	1.5	&	8.5	&	1790	&	6440	&	0.34	&	67	&	$0.5|0.0|0.5$	&	1.5	&	5.8	&	1580	&	2300	&	0.26	\\	
4	&	$1.0|0.0|0.0$	&	2.0	&	9.9	&	3380	&	9710	&	0.37	&	68	&	$0.5|0.0|0.5$	&	2.0	&	6.7	&	2860	&	3420	&	0.29	\\	
5	&	$1.0|0.0|0.0$	&	3.0	&	12.3	&	8330	&	17100	&	0.42	&	69	&	$0.5|0.0|0.5$	&	3.0	&	8.3	&	6830	&	6300	&	0.34	\\	
6	&	$1.0|0.0|0.0$	&	6.0	&	17.5	&	36500	&	43200	&	0.49	&	70	&	$0.5|0.0|0.5$	&	6.0	&	11.7	&	29800	&	16200	&	0.42	\\	
7	&	$1.0|0.0|0.0$	&	12.0	&	22.6	&	134000	&	85300	&	0.54	&	71	&	$0.5|0.0|0.5$	&	12.0	&	15.1	&	110000	&	33800	&	0.49	\\	
8	&	$1.0|0.0|0.0$	&	20.0	&	28.1	&	353000	&	156000	&	0.60	&	72	&	$0.5|0.0|0.5$	&	20.0	&	18.7	&	299000	&	61000	&	0.50	\\	
9	&	$0.0|1.0|0.0$	&	0.5	&	3.4	&	179	&	409	&	0.14	&	73	&	$0.3|0.0|0.7$	&	0.5	&	3.4	&	159	&	451	&	0.12	\\	
10	&	$0.0|1.0|0.0$	&	1.0	&	4.0	&	689	&	733	&	0.18	&	74	&	$0.3|0.0|0.7$	&	1.0	&	4.3	&	607	&	977	&	0.20	\\	
		 \hline

        \hline
        \vspace{0.05cm}
    \end{tabular}
     \item{\small Note: Reduced table shown, full data available as supplemental. }
\end{table*}

The addition of dipole and quadrupole components was explored in Paper I. We showed the fields to add in one hemisphere and subtract in the other. Similar to the dipole, the octupole component belongs in the ``primary'' symmetry family having anti-symmetric field polarity about the equator \citep{mcfadden1991reversals}. Addition of any primary geometries with any ``secondary'' family quadrupole (equatorially symmetric) would be expected to behave qualitatively similar. A different behaviour is expected from the addition of the two primary geometries (dipole-octupole). Here the field addition and subtraction is primarily governed by the relative orientations of the field with respect to one another. Aligned fields will combine constructively over the pole and subtract from one another in the equatorial region. Anti-aligned primary fields, conversely, will subtract on the pole and add over the equator. 

Including the results from Paper I, this work includes combinations of all the possible permutations of the axisymmetric dipole, quadrupole and octupole magnetic geometries. Table \ref{Parameters} contains a complete list of stellar parameters for the cases computed within this work. Parameters for the dipole-quadrupole combined field cases are available in Table 1 of Paper I. It is noted that in the course of the current work, the pure dipolar and quadrupole cases are re-simulated, see Table \ref{Parameters}.

\subsection{Derived Stellar Wind Properties}
The simulations produce steady state solutions for density, $\rho$, pressure, $p$, velocity, $\bf v$, and magnetic field strength, $\bf B$, for each stellar wind case. From these results, the behaviour of the spin down torque is ascertained. 
The torque on the star, $\tau$, due to the loss of angular momentum in the stellar wind is calculated,
\begin{equation}
\tau=\int_{\text{A}}\Lambda\rho{\bf v} \cdot d{\bf A},
\end{equation}
where the angular momentum flux, given by ${\bf F_{\text{AM}}}=\Lambda\rho{\bf v}$ \citep{keppens2000stellar}, is integrated over spherical shells of area $A$ (outside the closed field regions). $\Lambda$ is given by,
\begin{equation}
\Lambda(r,\theta)=rsin\theta\bigg(v_{\phi}-\frac{B_{\phi}}{\rho}\frac{|{\bf B_p}|^2}{{\bf v_p \cdot B_p}}\bigg).
\end{equation}
Similarly, the mass loss rate from our wind solutions is calculated,
\begin{equation}
\dot{M}=\int_{\text{A}}\rho{\bf v} \cdot d{\bf A}.
\end{equation}
An average Alfv\'en radius is then defined, in terms of the torque, mass loss rate, $\dot{M}$ and rotation rate, $\Omega_*$,
\begin{equation}
\langle R_{\text{A}}\rangle\equiv\sqrt{\frac{\tau}{\dot{M}\Omega_*}},
\label{averageAlfven}
\end{equation} 
In this formulation, $\langle R_{\text{A}}\rangle/R_*$ is defined as a dimensionless efficiency factor, by which the magnetised wind carries angular momentum from the star, i.e. a larger average Alfv\'en radius produces a larger torque for a fixed rotation rate and mass loss rate,
\begin{equation}
\tau=\dot{M}\Omega_*R_*^2\bigg(\frac{\langle R_{\text{A}}\rangle}{R_*}\bigg)^2.
\label{torque}
\end{equation}
In ideal MHD, $\langle R_{\text{A}}\rangle$ is associated with a cylindrical Alfv\'en radius, which acts like a ``lever arm'' for the spin-down torque on the star.

\begin{figure*}
    \includegraphics[width=\textwidth]{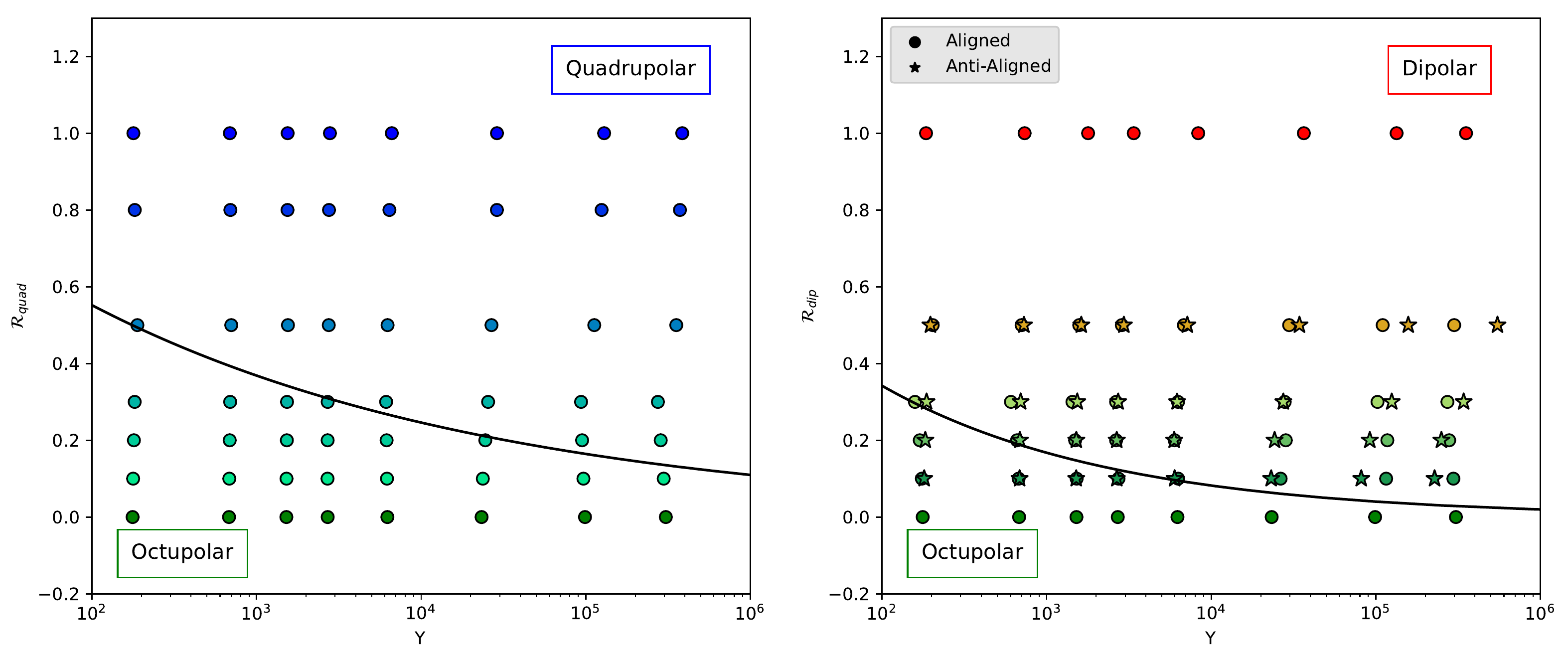}
     \caption{The two parameter spaces first examined in this work, quadrupole-octupole (left) and dipole-octupole (right), shown in terms of $\Upsilon$ and either $\mathcal{R}_{\text{quad}}$ or $\mathcal{R}_{\text{dip}}$ (equation \ref{rvalue}) respectively. Each point represents a simulation using the PLUTO code, with the colour of each point labelling them throughout this work, depending on their relative combination of field components. The black solid lines represent $\Upsilon_{crit}$ for each combination, where a change of regime from Alfv\'en radius scaling with as a pure octupole (bottom left) to scaling with the lowest order component, either dipole or quadrupole (upper right).}
     \label{Up_crit}
  \end{figure*}

The methodology of this work follows closely that of Paper I, in which we produce semi-analytic formulations for $\langle R_{\text{A}} \rangle$ in terms of the wind magnetisation, $\Upsilon$, as defined in previous works \citep{matt2008accretion, matt2012magnetic, reville2015effect, pantolmos2017magnetic},
\begin{equation}
\Upsilon=\frac{B_*^2R_*^2}{\dot{M}v_{\text{esc}}},
\label{upsilon}
\end{equation}
where $B_*$ is now the characteristic polar field; which is split amongst the different geometries using the ratios, $\mathcal{R}_{\text{dip}}$, $\mathcal{R}_{\text{quad}}$ and $\mathcal{R}_{\text{oct}}$. The values of $\Upsilon$ produced from the steady state solutions are indirectly controlled by increasing the value of $v_{\text{A}}/v_{\text{esc}}$. This increases the polar magnetic field strength for a given density normalisation. The mass loss rate is similarly uncontrolled and evolves to steady state, depending mostly on our choice of Parker wind parameters, but is also adjusted self-consistently by the magnetic field. The values of $\Upsilon$ are tabulated in Table \ref{Parameters}, along with $\mathcal{R}_l$ values, magnetic field strengths given by $v_{\text{A}}/v_{\text{esc}}$, and the average Alfv\'en radii for each case simulated. Results for combined dipole-quadrupole cases are available in Table 1 of Paper I. Figure~\ref{Up_crit} shows the parameter space of simulations with their value of $\Upsilon$ against the different ratios for either quadrupole-octupole or dipole-octupole cases, with the lower order geometry ratio labelling the cases ($\mathcal{R}_{\text{quad}}$ and $\mathcal{R}_{\text{dip}}$ respectively).

\section{Wind Solutions and $\langle R_{\text{A}}\rangle$ Scaling Relations}
\subsection{Single Geometry Winds}

\begin{table}
\caption{Single Component Fit Parameters to equation (\ref{single_mode})}
\label{fitValues}
\center
    \begin{tabular}{c|cc}
        \hline\hline
        Topology($l)$ & $K_{\text{s}}$   & $m_{\text{s}}$    \\ \hline
        Dipole ($1$)    &  $1.53\pm0.03$  & $0.229\pm0.002$               \\ 
        Quadrupole ($2$)    &  $1.70\pm0.02$  & $0.134\pm0.002$           \\   
        Octupole ($3$)    &  $1.80\pm0.01$  & $0.087\pm0.001$           \\   
        \hline
    \end{tabular}
        \item{\small Note: Fit values deviate slightly from those presented in Paper I due to the more accurate numerical results found with logarithmic grid spacing, used here. }
\end{table}

   \begin{figure}
    \includegraphics[width=0.47\textwidth]{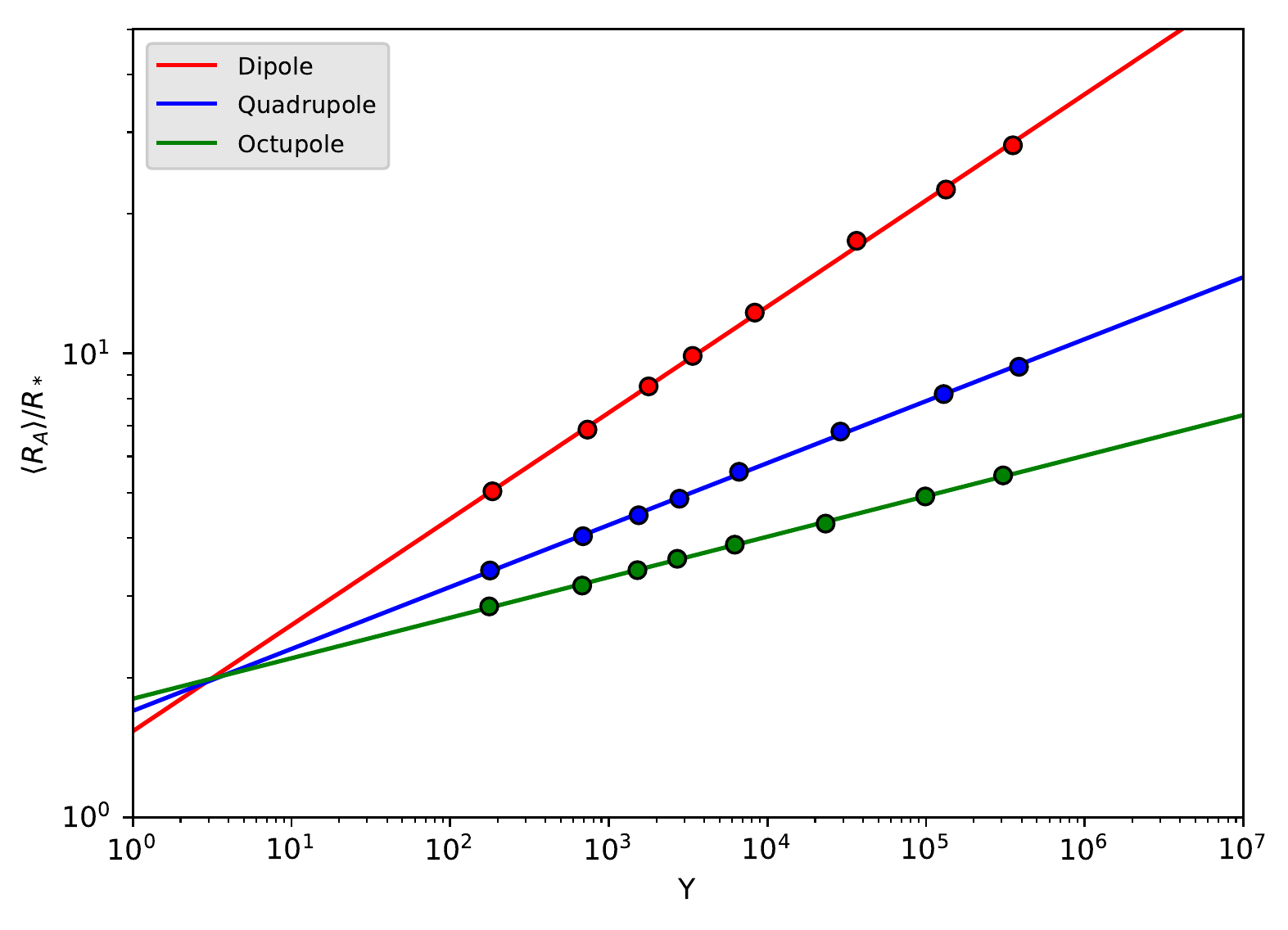}
     \caption{Average Alfv\'en radius vs the wind magnetisation, $\Upsilon$ (equation \ref{upsilon}) in our simulations with single geometries (points). Different scaling relations are shown for each pure geometry (solid lines). Higher $l$ order geometries produce a smaller Alfv\'en radius and thus smaller spin-down torque for a given polar field strength and mass loss rate. A similar result was first shown by \cite{reville2015effect}.}
     \label{Upsilon_puremodes}
  \end{figure}

For single magnetic geometries, increasing the complexity of the field decreases the effectiveness of the magnetic braking process by reducing the average Alfv\'en radius (braking lever arm) for a given field strength \citep{garraffo2015dependence}. The impact of changing field geometries on the scaling of the Alfv\'en radius for thermally driven winds was shown by \cite{reville2015effect} for the dipole, quadrupole and octupole geometries. We repeat the result of \cite{reville2015effect} for a slightly hotter coronal temperature wind, $c_{\text{s}}=0.25$ in our cases, compared to $c_{\text{s}}=0.222$. This temperature more reasonably approximates the solar wind terminal velocity, typically resulting in a wind speed of $\approx500$km/s at 1AU for solar parameters. For each magnetic geometry, we simulate 8 different field strengths changing the input value of $v_{\text{A}}/v_{\text{esc}}$ as tabulated in Table \ref{Parameters} (cases 1-24). 

Each wind solution gives a value for the Alfv\'en radius, $\langle R_{\text{A}} \rangle$, and the wind magnetisation, $\Upsilon$. These values are represented in Figure~\ref{Upsilon_puremodes} as coloured dots, and their scaling can be described using the Alfv\'en radius relation from \cite{matt2008accretion}, with three precise power law relations for the different magnetic geometries, as found previously in the work of \cite{reville2015effect}.
\begin{equation}
\frac{\langle R_{\text{A}} \rangle}{R_*}=K_{\text{s}}\Upsilon^{m_{\text{s}}},
\label{single_mode}
\end{equation}
where $K_{\text{s}}$ and $m_{\text{s}}$ are fit parameters for this relation, which utilises the surface field strength. Best fit parameters for each geometry tabulated in Table \ref{fitValues}. 

With increasing $l$ values, the higher order geometries produce increasingly shallow slopes with wind magnetisation, such that they approach a purely hydrodynamical lever arm i.e. the wind carries away angular momentum corresponding to the surface rotation alone, with the torque efficiency equal to the average cylindrical radius of the stellar surface from the rotation axis, $\langle R_{\text{A}} \rangle/R_*=(2/3)^{1/2}$ \citep{mestel1968magnetic}. Any significant magnetic braking in sun-like stars will therefore be predominantly mediated by the lowest order components. 
  
\subsection{Combined Magnetic Geometries}
Based on work performed in Paper I, we anticipate the behaviour of the average Alfv\'en radius for magnetic field geometries which contain, dipole, quadrupole and octupole components. The dipole component, having the slowest radial decay, is expected to govern the field strength at large distances, then the field should scale like the quadrupole at intermediate distances and finally, close to the star, the field should scale like the octupole geometry. The Alfv\'en radius formulation therefore takes the form of a twice broken power law,
\begin{equation}
  \frac{\langle R_{\text{A}} \rangle}{R_*}=\max\Bigg\{
  \begin{array}{@{}ll@{}}
    K_{\text{s,dip}}[\mathcal{R}_{\text{dip}}^2\Upsilon]^{m_{\text{s,dip}}},  \\
    K_{\text{s,quad}}[(|\mathcal{R}_{\text{dip}}|+|\mathcal{R}_{\text{quad}}|)^2\Upsilon]^{m_{\text{s,quad}}}, \\
    K_{\text{s,oct}}[(|\mathcal{R}_{\text{dip}}|+|\mathcal{R}_{\text{quad}}|+|\mathcal{R}_{\text{oct}}|)^2\Upsilon]^{m_{\text{s,oct}}},
  \end{array}
  \label{DQO_law}
\end{equation} 
which approximates the simulated values of the average Alfv\'en radius. Note $|\mathcal{R}_{\text{dip}}|+|\mathcal{R}_{\text{quad}}|+|\mathcal{R}_{\text{oct}}|=1$, such that the final scaling depends purely on the total $\Upsilon$.

Here we present simulation results from combinations of each field, sampling a range of mixing fractions and field strengths. These are used to validate this semi-analytic prescription for predicting the spin-down torque on a star, due to a given combination of axisymmetric magnetic fields. 

\subsubsection{Dipole Combined with Quadrupole}
\begin{figure}
    \includegraphics[width=0.5\textwidth]{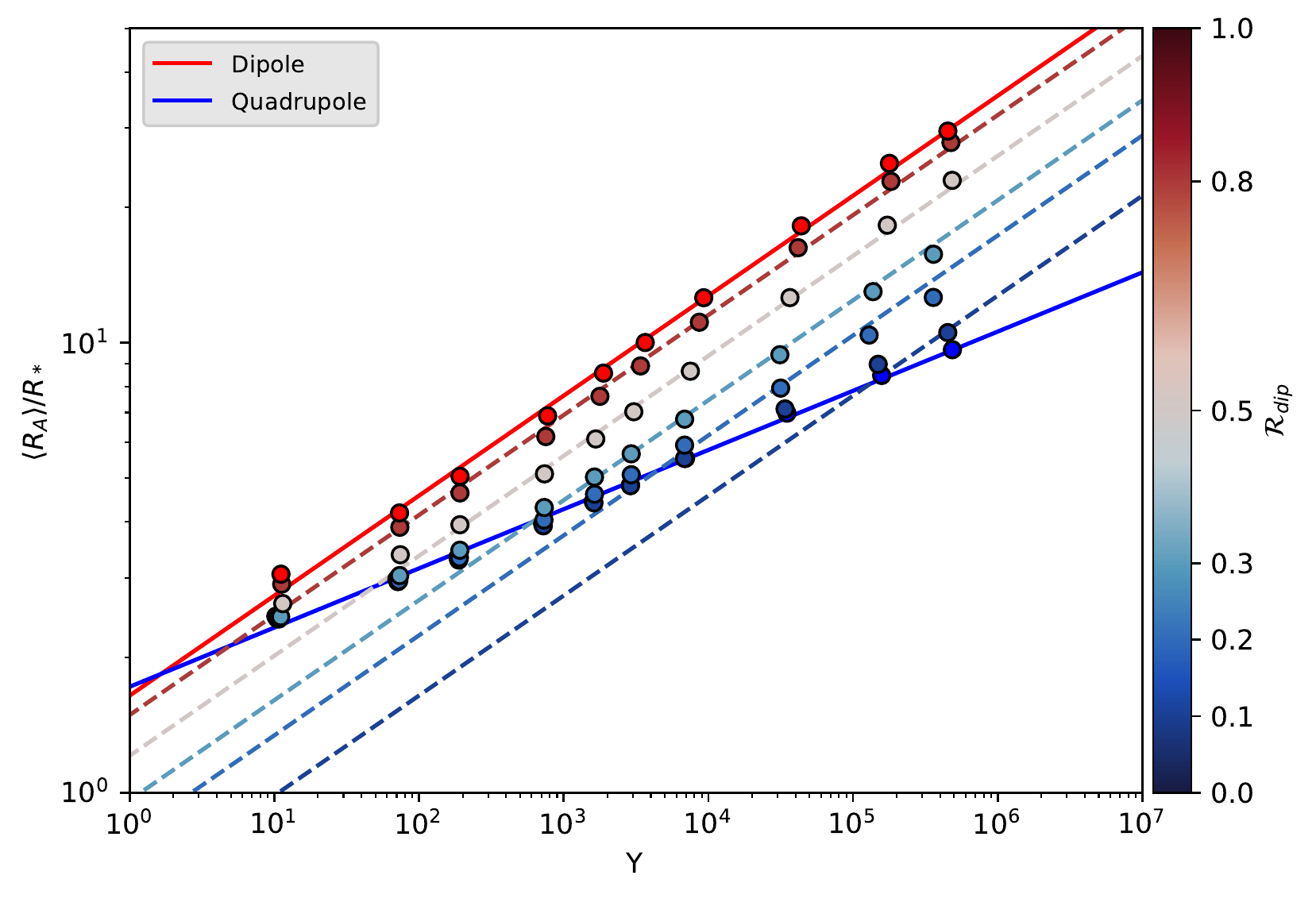}
     \caption{Average Alfv\'en radius scaling with wind magnetisation, $\Upsilon$, for the different combinations of dipole and quadrupole, from the study in Paper I (points). Solid lines show scaling for pure dipole and quadrupole. The deviation from single power laws shows how the combination of dipole and quadrupole fields modifies the Alfv\'en radius scaling, compared to single geometries. The scaling predicted by only considering the fractional dipole component is plotted with multiple dashed coloured lines corresponding to the different $\mathcal{R}_{\text{dip}}$ values. This shows that $\langle R_{\text{A}}\rangle/R_*$ scales with the dipole component only, unless the quadrupole is dominant at a distance of $\approx R_{\text{A}}$.}
     \label{Upsilon_DQ}
  \end{figure}

The regime of dipole and quadrupole combined geometries is presented in Paper I. We briefly reiterate the results here displaying values from that study in Figure~\ref{Upsilon_DQ}. 

These fields belong to different symmetry families, primary and secondary. As such their addition creates a globally asymmetric field about the equator, with the north pole in this case being stronger than the south. The relative fraction of the two components alters the location of the current sheet/streamers, which appear to resemble the dominant global geometry. 

   \begin{figure*}
   \centering
    \includegraphics[width=0.9\textwidth]{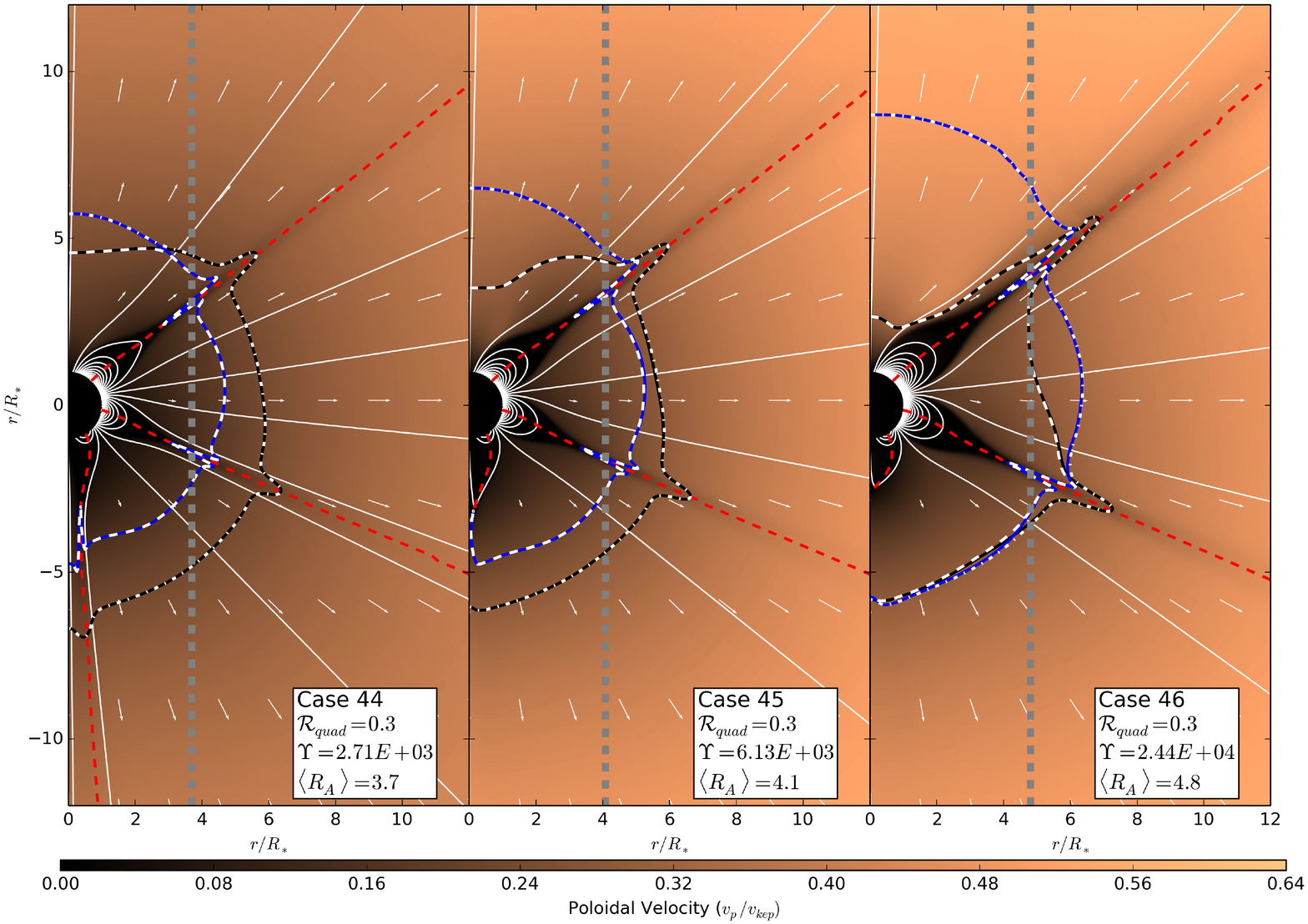}
     \caption{Steady state solutions for the quadrupole-octupole combined geometry cases 44, 45 and 46, showing a progression from weaker to stronger magnetisation ($\Upsilon$) from left to right. The colour background represents the poloidal speed normalised by the Keplerian speed (e.g. $\approx400$km/s for Solar parameters). Deadzones are therefore in black. Thin white lines trace the magnetic field, with red dashed lines highlighting the field polarity reversals (i.e. where $B_r=0$). Alfv\'en and sonic surfaces are indicated with thick blue and black lines respectively, with the fast and slow magnetosonic surfaces represented as dot-dash and dashed white lines. Vertical grey dashed lines show the average Alfv\'en radius $\langle R_{\text{A}}\rangle$ (equation \ref{averageAlfven}), representing the torque efficiency, scales with the size of the Alfv\'en surface. The asymmetry of the global magnetic field about the equator is shown, with a qualitatively similar behaviour to the dipole-quadrupole simulations in Paper I.}
     \label{QO_Example}
  \end{figure*}

It is shown in Paper I that the quadrupole component has a faster radial decay than the dipole, and therefore at large distances only the dipole component of the field influences the location of the Alfv\'en radius. Closer to the star, the total field decays radially like the quadrupole, with the dipole component adding its strength, so near to the star the Alfv\'en radius scaling depends on the total field strength. Therefore we developed a broken power law to describe the behaviour of the average Alfv\'en radius scaling with wind magnetisation, which uses the maximum of either the quadrupole slope using the total field strength, as if the combined field decays like a quadrupole, (solid blue line) or the dipolar slope using only the dipole component (shown in colour-coded dashed lines). The dipole component of the wind magnetisation is formulated as,
\begin{equation}
\Upsilon_{\text{dip}}=\bigg(\frac{B_*^{l=1}}{B_*}\bigg)^2\frac{B_*^2R_*^2}{\dot{M}v_{\text{esc}}}=\mathcal{R}_{\text{dip}}^2\Upsilon.
\label{upsilon_dipole}
\end{equation}
Mathematically, equation (\ref{DQO_law}) becomes the broken power law from Paper I when $\mathcal{R}_{\text{oct}}=0$,
\begin{equation}
  \frac{\langle R_{\text{A}} \rangle}{R_*}=\left\{
  \begin{array}{@{}ll@{}}
    K_{\text{s,dip}}[\mathcal{R}_{\text{dip}}^2\Upsilon]^{m_{\text{s,dip}}}, & \text{if}\ \Upsilon>\Upsilon_{crit}(\mathcal{R}_{\text{dip}}), \\
    K_{\text{s,quad}}[\Upsilon]^{m_{\text{s,quad}}}, & \text{if}\ \Upsilon\leq\Upsilon_{crit}(\mathcal{R}_{\text{dip}}),
  \end{array}\right.
\end{equation} 
where the octupolar relation is ignored, and $|\mathcal{R}_{\text{dip}}|+|\mathcal{R}_{\text{quad}}|=1$. Here $\Upsilon_{crit}$ describes the intercept of the dipole component and quadrupole slopes,
\begin{equation}
\Upsilon_{crit}(\mathcal{R}_{\text{dip}})=\bigg[\frac{K_{\text{s,dip}}}{K_{\text{s,quad}}}\mathcal{R}_{\text{dip}}^{2m_{\text{s,dip}}} \bigg]^{\frac{1}{m_{\text{s,quad}}-m_{\text{s,dip}}}}.
\end{equation}

Equation (\ref{DQO_law}) further expands the reasoning above to include any field combination of the axisymmetric dipole, quadrupole and octupole. The following sections test this formulation against simulated combined geometry winds.

\subsubsection{Quadrupole Combined with Octupole}
Stellar magnetic fields containing both a quadrupole and octupole field component present another example of primary and secondary family fields in combination.  As with the axisymmetric dipole-quadrupole addition, the relative orientation of the two components simply determines which regions of magnetic field experience addition and subtraction about the equator, so that the torque and mass loss rate do not depend on their relative orientation. Compared with the dipole component, both fields are less effective in generating a magnetic lever arm to brake rotation at a given value of $\Upsilon$. 

We test the validity of equation (\ref{DQO_law}), setting $\mathcal{R}_{\text{dip}}=0$, and systematically varying the value of $\mathcal{R}_{\text{quad}}$, with the octupole fraction comprising the remaining field, $\mathcal{R}_{\text{oct}}=1-\mathcal{R}_{\text{quad}}$. Five mixed case values are selected ($\mathcal{R}_{\text{quad}}=0.8, 0.5, 0.3, 0.2, 0.1$) that parametrise the mixing of the two geometries. Steady state wind solutions are displayed in Figure~\ref{QO_Example}, showing, as with dipole-quadrupole addition, the equatorially asymmetric fields produced. With increasing polar field strength, the streamers are observed shift towards the lowest order geometry morphology (quadrupolar in this case), as was shown for the dipole in Paper I.

\begin{figure}
    \includegraphics[width=0.5\textwidth]{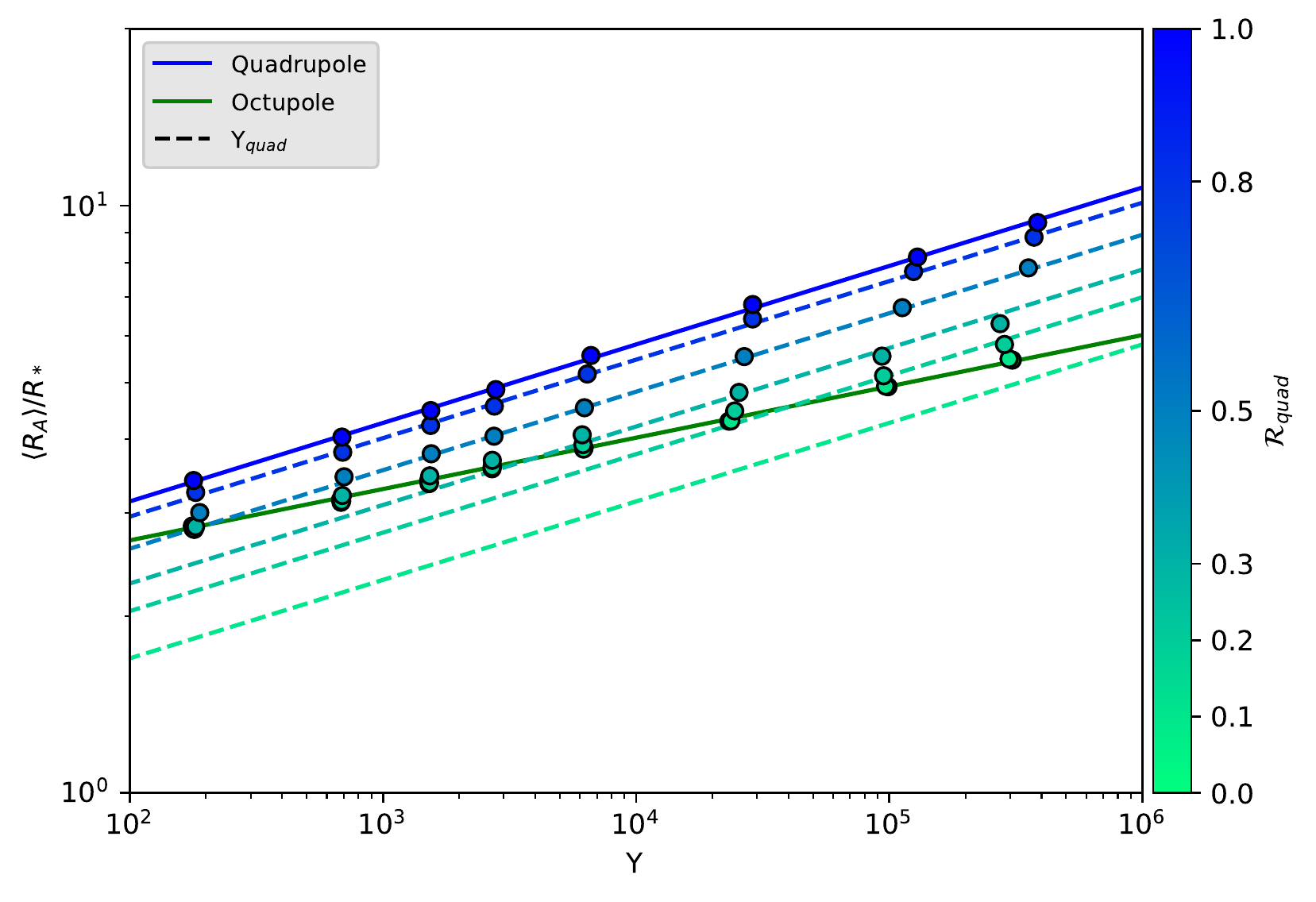}
     \caption{Average Alfv\'en radius vs wind magnetisation, $\Upsilon$, for the different combinations of quadrupole and octupole, in a similar format to Figure~\ref{Upsilon_DQ}. Colour-coded dashed lines relate to the prediction considering only the quadrupolar component of the field for each $\mathcal{R}_{\text{quad}}$. The combinations shown here behave in a similar manner to dipole-quadrupole combined fields, in a sense that the lower order field (with the lowest $l$) governs the Alfv\'en radius for large wind magnetisations, $\Upsilon$, and the higher order (large $l$) controlling the low magnetisation scaling. }
     \label{Upsilon_QO}
  \end{figure}

The average Alfv\'en radii and wind magnetisation are shown in Figure~\ref{Upsilon_QO}. The behaviour of $\langle R_{\text{A}} \rangle$ is quantitatively similar to that of the dipole-quadrupole addition, where combined field cases are scattered between the two pure geometry scaling relations. The range of available $\langle R_{\text{A}} \rangle$ values between the pure quadrupole and octupole scaling relations (solid blue and green respectively) is reduced compared to the previous dipole-quadrupole, due to the weaker dependence of the Alfv\'en radius with wind magnetisation. 

   \begin{figure}
    \includegraphics[width=0.5\textwidth]{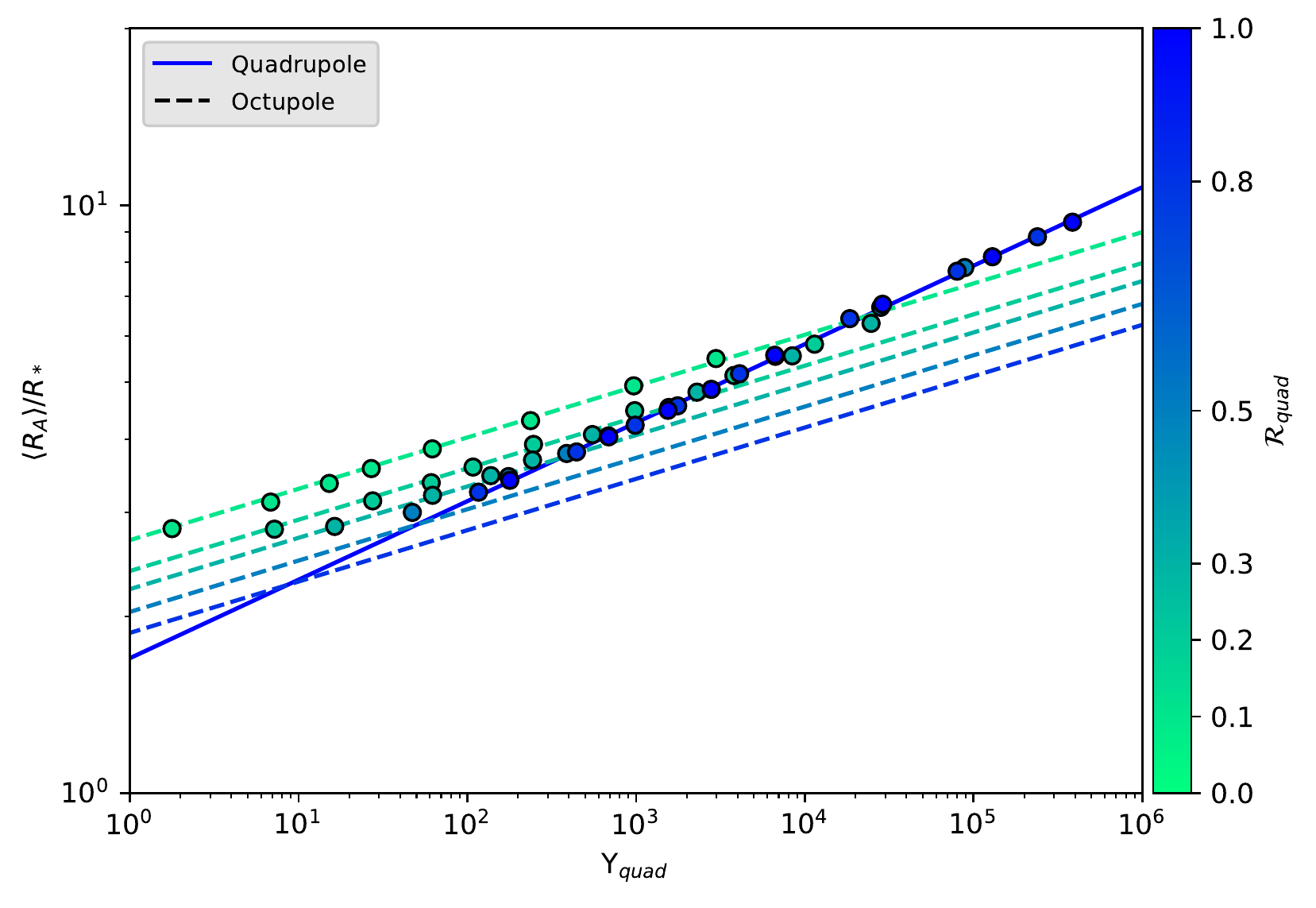}
     \caption{Average Alfv\'en radius vs the quadrupolar component of the wind magnetisation, $\Upsilon_{\text{quad}}$, for cases with mixed quadrupole and octupole components (points). The solid blue line shows the prediction based on the quadrupole component only (equation \ref{up_quad_rel}). The dashed lines show the octupolar scaling (equation \ref{oct_scaling_re}). A broken power law composed of the quadrupolar component and the octupolar scaling ($\mathcal{R}_{\text{quad}}$ dependent) can be constructed similarly to work done in Paper I. The quadrupolar geometry dominates the scaling, for all of the $\mathcal{R}_{\text{quad}}$ values simulated here, at $\langle R_{\text{A}}\rangle/R_*\approx 9$. The point at which the quadrupolar geometry dominates for a given $\mathcal{R}_{\text{quad}}$ value can be approximated by considering the strength of the two fields at the Alfv\'en radii, i.e. the radial distance when the strength of the quadrupole matches or exceeds that of the octupole $B_{\text{quad}}/B_{\text{oct}} =\mathcal{R}_{\text{quad}}/(1-\mathcal{R}_{\text{quad}})(r/R_*)$.}
     \label{Up_quad_QO}
  \end{figure}
 
As required by equation (\ref{DQO_law}), with no dipolar component, we introduce the quadrupole component of $\Upsilon$ as,
\begin{equation}
\Upsilon_{\text{quad}}=\bigg(\frac{B_*^{l=2}}{B_*}\bigg)^2\frac{B_*^2R_*^2}{\dot{M}v_{\text{esc}}}=\mathcal{R}_{\text{quad}}^2\Upsilon,
\end{equation}
and the second relation in equation (\ref{DQO_law}) takes the form,
\begin{equation}
\frac{\langle R_{\text{A}} \rangle}{R_*}=K_{\text{s,quad}}[\Upsilon_{\text{quad}}]^{m_{\text{s,quad}}},
\label{up_quad_rel}
\end{equation}
where, $K_{\text{s,quad}}$ and $m_{\text{s,quad}}$ are determined from the pure geometry scaling, see Table \ref{fitValues}. 

The quadrupole component of the wind magnetisation is plotted for different $\mathcal{R}_{\text{quad}}$ values in Figure~\ref{Upsilon_QO}, showing an identical behaviour to the dipole component in the dipole-quadrupole combined fields. The $\Upsilon_{\text{quad}}$ formulation is shown within Figure~\ref{Up_quad_QO}, with the solid blue line described by equation (\ref{up_quad_rel}). This agrees with a large proportion of the wind solutions, with deviations due to a switch of regime onto the octupole relation, the third relation in equation (\ref{DQO_law}), 
\begin{equation}
\frac{\langle R_{\text{A}} \rangle}{R_*}=K_{\text{s,oct}}[\Upsilon]^{m_{\text{s,oct}}}=\frac{K_{\text{s,oct}}}{\mathcal{R}_{\text{quad}}^{2m_{\text{s,oct}}}}[\Upsilon_{\text{quad}}]^{m_{\text{s,oct}}},
\label{oct_scaling_re}
\end{equation}
shown with a solid green line in Figure~\ref{Upsilon_QO} and dashed colour-coded lines in Figure~\ref{Up_quad_QO}. As with the dipole-quadrupole addition, a broken power law can be formulated taking the maximum of either the octupole scaling or the quadrupole component scaling, for a given $\mathcal{R}_{\text{quad}}$ value. For the cases simulated, we find a deviation from this broken power law of no greater than $5\%$, with most cases showing a closer agreement. 

   \begin{figure*}
   \centering
    \includegraphics[width=0.9\textwidth]{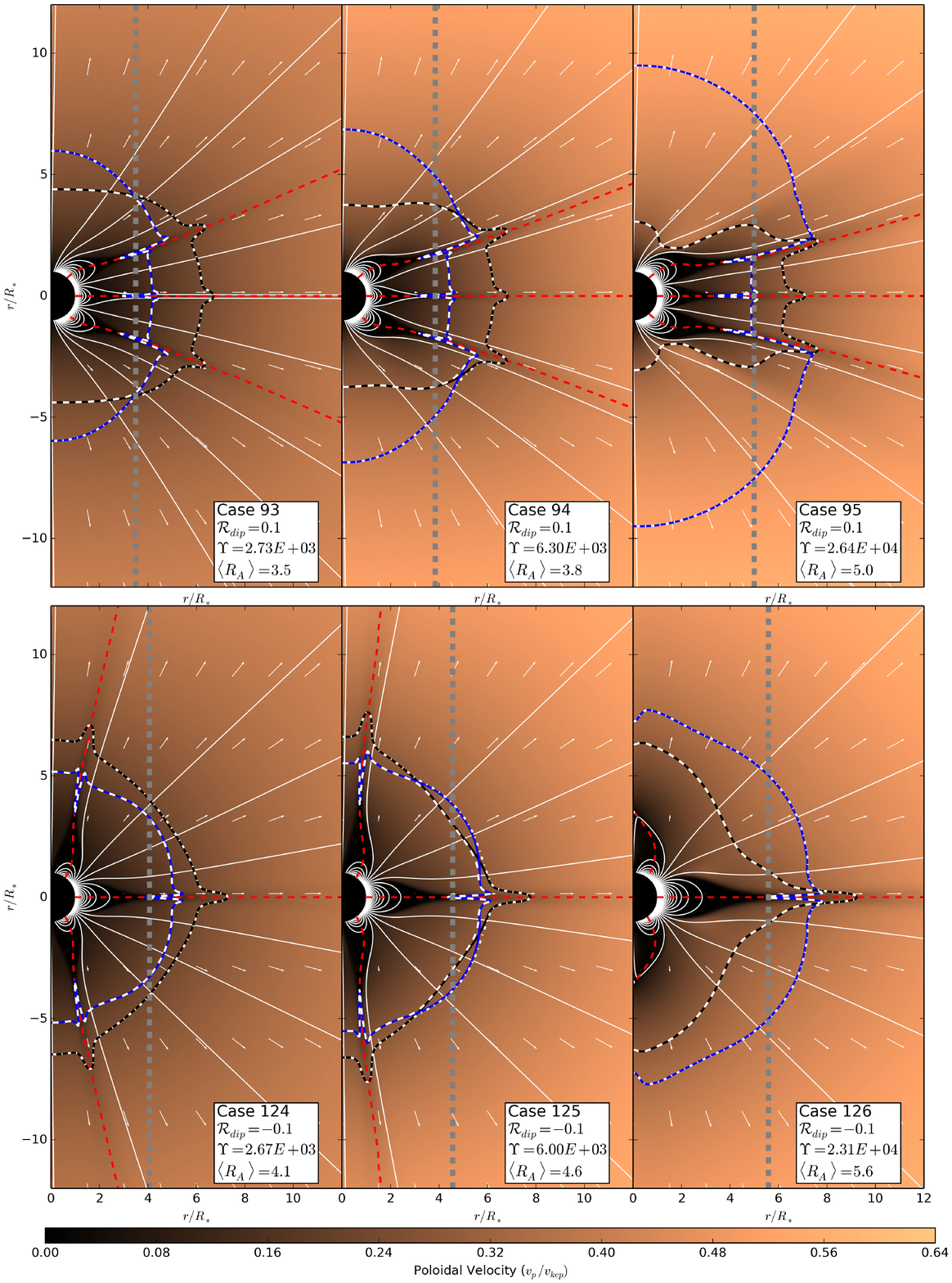}
     \caption{Steady state solutions for the dipole-octupole combined geometries with aligned fields (top row, cases 93, 94 and 95) and anti-aligned fields (bottom row, cases 124, 125 and 126). The format and lines are the same as Figure~\ref{QO_Example}. The aligned cases have field adding near the poles and subtracting near the equator, where the opposite is true for the anti-aligned cases. The difference in how these two cases combine results in a different shape of the Alfv\'en surface. Also, for the same magnetisation ($\Upsilon$), the anti-aligned cases, in general, systematically produce a larger torque efficiency ($\langle R_{\text{A}}\rangle$, vertical dashed grey lines). This is due to the these cases having a stronger field at low latitudes, where the angular momentum loss is more efficient.}
     \label{DO_Example}
  \end{figure*}

\subsubsection{Dipole Combined with Octupole}
Unlike the previous field combinations, both the dipole and octupole belong to the primary symmetry family and thus their addition produces two distinct field topologies for aligned or anti-aligned fields. Again, we test equation (\ref{DQO_law}), now with $\mathcal{R}_{\text{quad}}=0$. The field combinations are parametrised using the ratio of dipolar field to total field strength, $\mathcal{R}_{\text{dip}}$, with the remaining field in the octupolar component $\mathcal{R}_{\text{oct}}=1-\mathcal{R}_{\text{dip}}$. The ratio of dipolar field is varied ($\mathcal{R}_{\text{dip}}=0.5, 0.3, 0.2, 0.1$). Additionally we repeat these ratios for both aligned and anti-aligned fields. This produces eight distinct field geometries that cover a range of mixed dipole-octupole fields. 

Figure~\ref{DO_Example} displays the behaviour of both aligned and anti-aligned cases with increasing field strength. The combination of dipolar and octupolar fields produces a complex field topology which is alignment dependent and impacts the local flow properties of the stellar wind. The symmetric property of the global field is maintained about the equator. Aligned combinations have magnetic field addition over the poles which increases the Alfv\'en speed, producing a larger Alfv\'en radius over the poles. However, the fields subtract over the equator which reduces the size of the Alfv\'en radius over the equator; top panel of Figure~\ref{QO_Example}. The bottom panel shows anti-aligned mixed cases to exhibit the opposite behaviour, with a larger equatorial Alfv\'en radius and a reduction to the size of the Alfv\'en surface at higher latitudes. The torque averaged Alfv\'en radius is shown by the grey dashed lines in each case, representing the cylindrical Alfv\'en radius $\langle R_{\text{A}} \rangle$. For the simulations in this work, the anti-aligned cases produce a larger lever arm compared with their aligned counterparts, with a few exceptions. In general, the increased Alfv\'en radius at the equator for the anti-aligned fields is more effective at increasing the torque averaged Alfv\'en radius compared with the larger high-latitude Alfv\'en radius in the aligned fields cases.

The location of the current sheets are shown in Figure~\ref{DO_Example} using red dashed lines. As noted with the dipole-quadrupole addition in Paper I, the global dipolar geometry is restored with increasing fractions of the dipole component or increased field strength for a given mixed geometry. The latter is shown in Figure~\ref{DO_Example} for both aligned and anti-aligned cases. With increased field strength, a single dipolar streamer begins to be recovered over the equator. A key difference between the two field alignments is the asymptotic location of the three streamers. In the case of an aligned octupole component, increasing the total field strength for a given ratio forces the streamers towards the equator at which point they begin to merge into the dipolar streamer. With an anti-aligned octupole component, the opposite is found, with the high latitude streamers forced towards the poles and onto the rotation axis. It is unclear if this effect is significant itself on influencing the global torque.

\begin{figure}
    \includegraphics[width=0.5\textwidth]{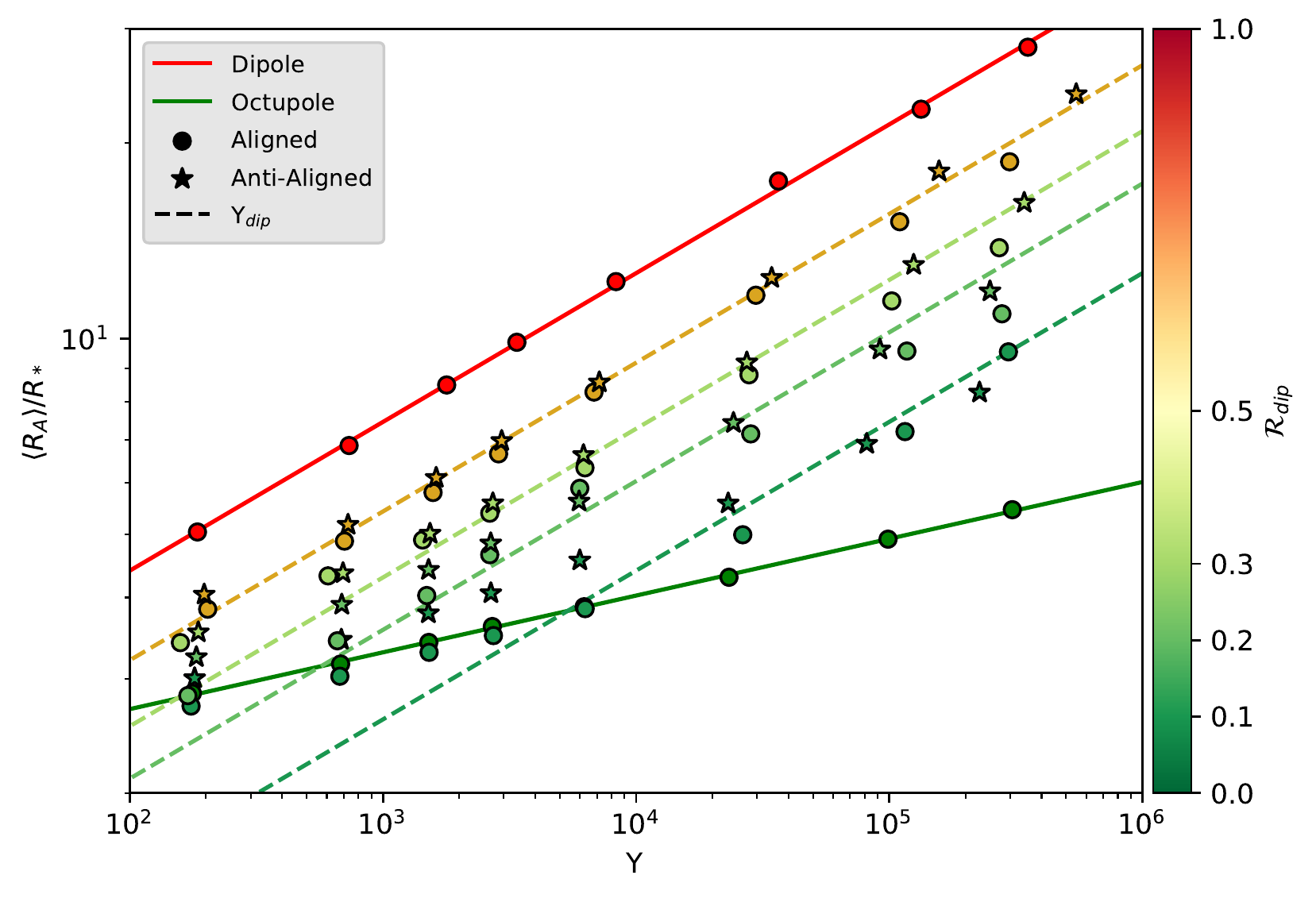}
     \caption{Average Alfv\'en radius scaling with wind magnetisation, $\Upsilon$, for the different combinations of dipole and octupole. The fields are either added aligned at the poles (points) or anti-aligned (stars). Dashed lines show the dipole component scaling, colour-coded to match the simulated values of $\mathcal{R}_{\text{dip}}$. The overall behaviour here is similar to the previous mixed combined fields, with the lower order field governing the Alfv\'en radius for large wind magnetisations. However the different field alignments appear to scatter around the $\Upsilon_{\text{dip}}$ approximation, with the anti-aligned cases typically having larger $R_{\text{A}}$ than the aligned cases, for the same $\Upsilon$.}
     \label{Upsilon_DO}
  \end{figure}

\begin{figure}
    \includegraphics[width=0.5\textwidth]{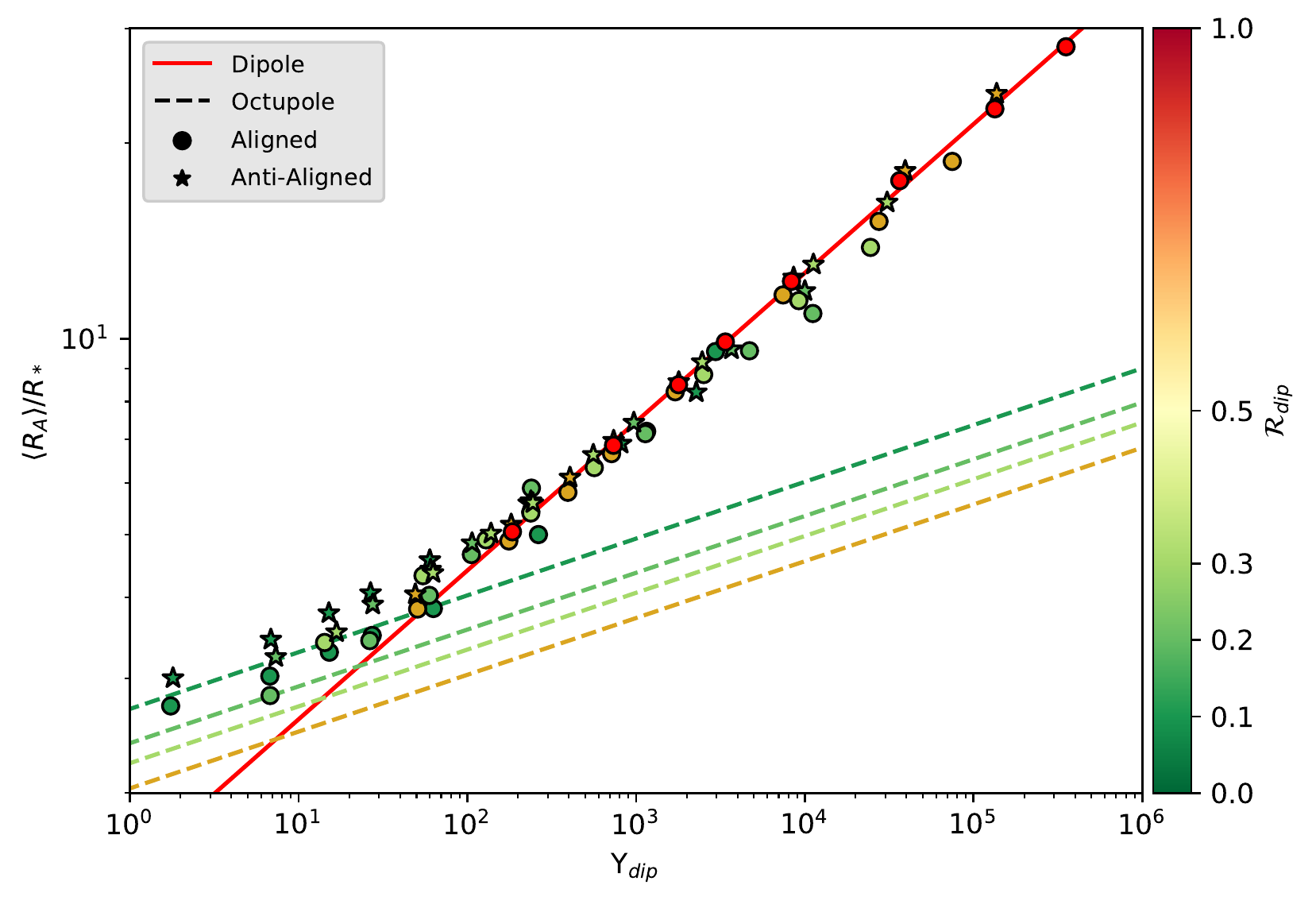}
     \caption{Average Alfv\'en radius scaling with only the dipolar component of the wind magnetisation, $\Upsilon_{\text{dip}}$, for cases with combined dipole and octupole components. Aligned field are shown with circles, anti-aligned with stars. The parameter space investigated here is well approximated by the dipole component scaling relation (solid red line). Generally the aligned field cases are shown to under-shoot the dipole component approximation whilst the anti-aligned cases match the power law with similar agreement as the previous combined geometries. The qualitative behaviour is again similar to the previous combined cases, however due to the larger difference in radial decay of the field i.e. $B_{\text{dip}}/B_{\text{oct}} =\mathcal{R}_{\text{dip}}/(1-\mathcal{R}_{\text{dip}})(r/R_*)^2$, the dipole dominates at much smaller $R_{\text{A}}\approx3$.}
     \label{Up_dip_DO}
  \end{figure}

Using equation (\ref{DQO_law}), with no quadrupolar component, we anticipate the dipolar component (first relation) will be the most significant in governing the global torque. Figures \ref{Upsilon_DO} and \ref{Up_dip_DO} show the dipole-octupole cases following the expected behaviour, as observed for dipole-quadrupole and quadrupole-octupole combinations. We see that the average Alfv\'en radius either follows the dipole component scaling ($\Upsilon_{\text{dip}}$), or the octupole scaling relation, 
\begin{equation}
\frac{\langle R_{\text{A}} \rangle}{R_*}=K_{\text{s,oct}}[\Upsilon]^{m_{\text{s,oct}}}=\frac{K_{\text{s,oct}}}{\mathcal{R}_{\text{dip}}^{2m_{\text{s,oct}}}}[\Upsilon_{\text{dip}}]^{m_{\text{s,oct}}}.
\label{oct_scaling_re2}
\end{equation}
However, as evident in both figures, there is a deviation from this scaling, with the strongest variations belonging to low $\mathcal{R}_{\text{dip}}$ cases. Anti-aligned cases follow the behaviour expected from Paper I with a much higher precision than the anti-aligned cases. Figure~\ref{Up_dip_DO} shows the dipole scaling to over-predict the aligned cases compared with the anti-aligned cases. This occurs as equation (\ref{DQO_law}) is a simplified picture of the actual dynamics within our simulations, and as such, it does not encapsulate all of the physical effects. The trends are still obvious for both aligned and anti-aligned cases, and the scatter simply represents a reduction to the precision of our formulation. 

Despite this deviation from predicted values, Figure~\ref{Up_dip_DO} shows the dipole component again to be the most significant in governing the global torque. With a more complex (higher $l$) secondary component, the dipole dominates the Alfv\'en radius scaling at a much lower wind magnetisation, when compared with the dipole-quadrupole combinations. For the dipole-octupole cases simulated, the dipole component dominates the majority of the simulated cases. For our dipole and octupole mixed fields the transition between regimes occurs at $\Upsilon_{\text{dip}}\approx100$, such that the $\langle R_{\text{A}} \rangle$ for fields with $\mathcal{R}_{\text{dip}}=0.1$, or higher, and a physically realistic wind magnetisation, will all be governed by the dipole component.

\subsubsection{Combined Dipole, Quadrupole and Octupole Fields}
In addition to the quadrupole-octupole and dipole-octupole combinations presented previously, we also perform a small set of simulations containing all three components. Their stellar wind parameters and results are tabulated in Table \ref{Parameters_extra}. We select a regime where the dipole does not dominate ($\mathcal{R}_{\text{dip}}=0.1$), to observe the interplay of the additional quadrupole and octupole components. We also utilise cases 89-96 and 121-128 from this work and cases 51-60 from Paper I, all of which sample varying fractions of quadrupole and octupole with a fixed $\mathcal{R}_{\text{dip}}=0.1$. These are compared against the three component cases, 129-160.

  \begin{table}
\caption{Input Parameters and Results from Simulations with Three Magnetic Components}
\label{Parameters_extra}
\center
\setlength{\tabcolsep}{1pt}
    \begin{tabular}{ccccccc}
        \hline\hline
Case	&	$\mathcal{R}_{\text{dip}}|\mathcal{R}_{\text{quad}}|\mathcal{R}_{\text{oct}}$	&	$v_{\text{A}}/v_{\text{esc}}$	&	$\langle R_{\text{A}}\rangle/R_*$	&	$\Upsilon$	&	$\Upsilon_{\text{open}}$	&	$\langle v(R_{\text{A}})\rangle/v_{\text{esc}} $	\\	\hline
129	&	$0.1|0.6|0.3$	&	0.5	&	3.1	&	181	&	289	&	1.09	\\	
130	&	$0.1|0.6|0.3$	&	1.0	&	3.6	&	698	&	502	&	1.33	\\	
131	&	$0.1|0.6|0.3$	&	1.5	&	4.0	&	1550	&	709	&	1.49	\\	
132	&	$0.1|0.6|0.3$	&	2.0	&	4.4	&	2760	&	923	&	1.61	\\	
133	&	$0.1|0.6|0.3$	&	3.0	&	4.9	&	6320	&	1400	&	1.81	\\	
134	&	$0.1|0.6|0.3$	&	6.0	&	6.3	&	27100	&	3030	&	2.17	\\	
135	&	$0.1|0.6|0.3$	&	12.0	&	7.9	&	111000	&	6430	&	2.65	\\	
136	&	$0.1|0.6|0.3$	&	20.0	&	9.3	&	308000	&	11200	&	3.09	\\	
137	&	$0.1|0.6|0.6$	&	0.5	&	2.7	&	182	&	194	&	0.97	\\	
138	&	$0.1|0.3|0.6$	&	1.0	&	3.1	&	702	&	326	&	1.17	\\	
139	&	$0.1|0.3|0.6$	&	1.5	&	3.4	&	1560	&	451	&	1.29	\\	
140	&	$0.1|0.3|0.6$	&	2.0	&	3.7	&	2760	&	585	&	1.37	\\	
141	&	$0.1|0.3|0.6$	&	3.0	&	4.2	&	6230	&	903	&	1.53	\\	
142	&	$0.1|0.3|0.6$	&	6.0	&	5.5	&	25600	&	2180	&	1.85	\\	
143	&	$0.1|0.3|0.6$	&	12.0	&	7.2	&	97000	&	4850	&	2.25	\\	
144	&	$0.1|0.3|0.6$	&	20.0	&	8.6	&	246000	&	8560	&	2.61	\\	
145	&	$0.1|0.6|-0.3$	&	0.5	&	3.2	&	34	&	312	&	1.13	\\	
146	&	$0.1|0.6|-0.3$	&	1.0	&	3.7	&	119	&	533	&	1.37	\\	
147	&	$0.1|0.6|-0.3$	&	1.5	&	4.1	&	258	&	765	&	1.53	\\	
148	&	$0.1|0.6|-0.3$	&	2.0	&	4.5	&	451	&	1000	&	1.65	\\	
149	&	$0.1|0.6|-0.3$	&	3.0	&	5.1	&	1020	&	1500	&	1.85	\\	
150	&	$0.1|0.6|-0.3$	&	6.0	&	6.5	&	4450	&	3400	&	2.21	\\	
151	&	$0.1|0.6|-0.3$	&	12.0	&	8.2	&	18600	&	7260	&	2.69	\\	
152	&	$0.1|0.6|-0.3$	&	20.0	&	10.1	&	55300	&	13200	&	3.17	\\	
153	&	$0.1|0.3|-0.6$	&	0.5	&	3.0	&	4	&	254	&	1.05	\\	
154	&	$0.1|0.3|-0.6$	&	1.0	&	3.5	&	21	&	430	&	1.25	\\	
155	&	$0.1|0.3|-0.6$	&	1.5	&	3.9	&	49	&	607	&	1.37	\\	
156	&	$0.1|0.3|-0.6$	&	2.0	&	4.2	&	91	&	782	&	1.49	\\	
157	&	$0.1|0.3|-0.6$	&	3.0	&	4.7	&	214	&	1160	&	1.65	\\	
158	&	$0.1|0.3|-0.6$	&	6.0	&	5.9	&	916	&	2440	&	2.01	\\	
159	&	$0.1|0.3|-0.6$	&	12.0	&	7.5	&	3770	&	5360	&	2.41	\\	
160	&	$0.1|0.3|-0.6$	&	20.0	&	9.3	&	11300	&	10200	&	2.85	\\		\hline
        \hline
        \vspace{0.05cm}
    \end{tabular}
\end{table}

   \begin{figure}
   \centering
    \includegraphics[width=0.45\textwidth]{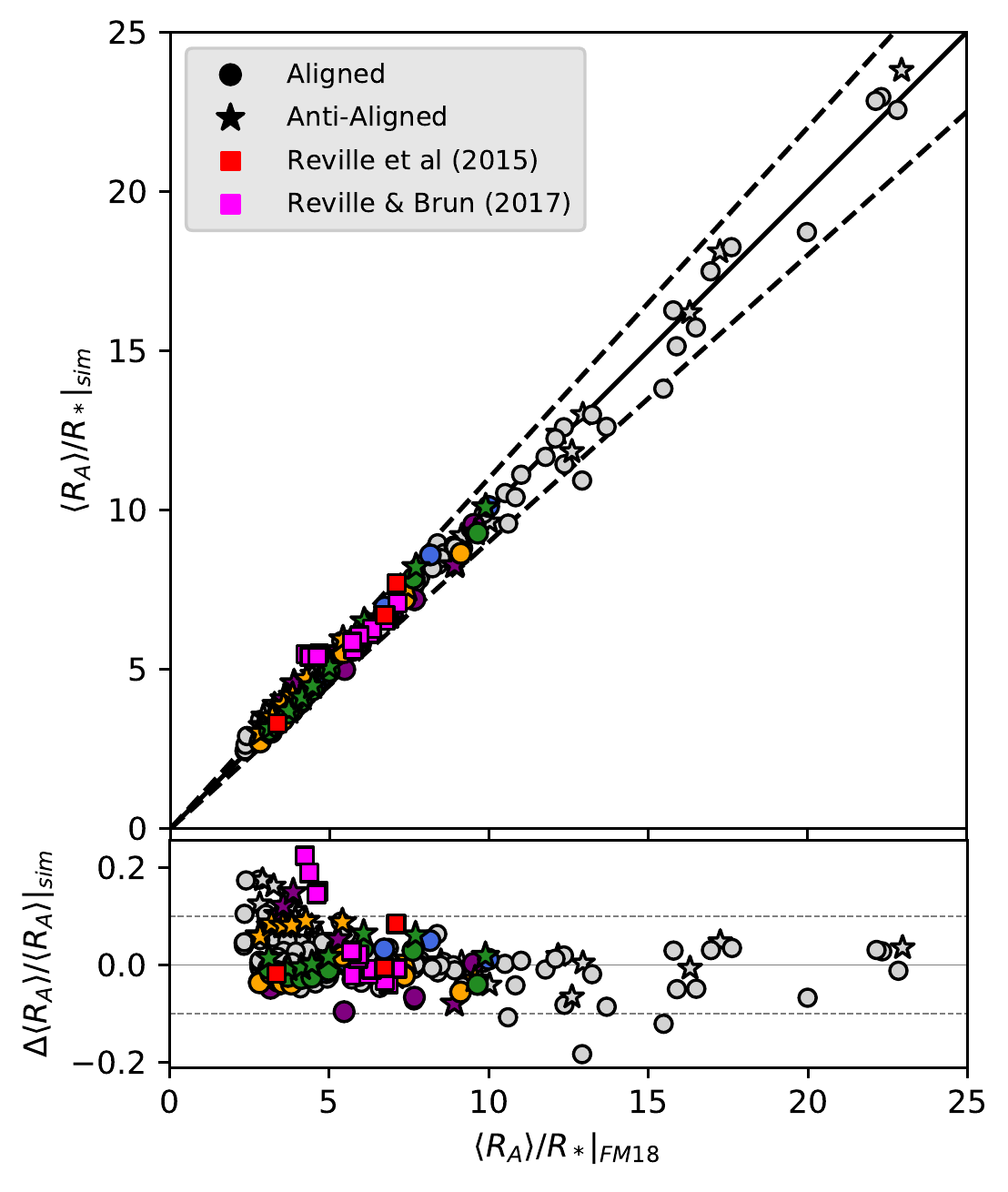}
     \caption{Comparison of the simulated Alfv\'en radii vs the predicted Alfv\'en radii using equation (\ref{DQO_law}), top panel. The line of agreement is shown with a solid black line, and the bounds of $10\%$ deviation from the predicted value are shown with black dashed lines. The bottom panel shows the residual, $(\langle R_{\text{A}}\rangle_{sim}-\langle R_{\text{A}}\rangle_{FM18})/\langle R_{\text{A}}\rangle_{sim}$, and the $10\%$ deviation with dashed lines. Cases 129-135 \& 145-152 are coloured purple and cases 137-144 \& 153-160 are coloured orange, different from the colour scheme of previous figures. The quadrupole and octupole dominated cases with $\mathcal{R}_{\text{dip}}$=0.1 are shown with their original colouring (blue and green respectively). All other simulations from this work, and Paper I, are shown in grey. Three red squares represent axisymmetric mixed field simulations from \cite{reville2015effect}. Thirteen magenta squares represent 3D non-axisymmetric simulations with $l_{max}=15$ from \cite{reville2017global} (the average Alfv\'en radius is computed differently than equation \ref{averageAlfven}).}
     \label{mixed_cases}
  \end{figure}

Equation (\ref{DQO_law}) is adopted, now using all three components, such that the results from these simulations are expected to scale in magnetisation like a twice broken power law. As noted with the dipole-octupole addition, the inclusion of an octupolar component introduces behaviours which will not be accounted for by this formulation, i.e. equation (\ref{DQO_law}) is independent of field alignments, etc. We aim to characterise this unaccounted for physics in terms of an available precision on the use of equation (\ref{DQO_law}). The simulated Alfv\'en radii are compared against their predicted values in Figure~\ref{mixed_cases}, along with the other simulations from this work (shown in white). The three component field combinations have a small dipolar component; therefore the dipolar scaling of the average Alfv\'en radius is rarely the dominant term in equation (\ref{DQO_law}). The different values of quadrupolar and octupolar field that comprise the remaining field strength govern the average Alfv\'en radius scaling for the majority of this parameter space. From Figure~\ref{mixed_cases}, the approximate formulation agrees well with the simulated values with the largest discrepancies emerging at smaller radii and for anti-aligned cases, see the residual plot below. A $10\%$ divergence from our prediction (dashed lines in both the top and bottom panel of Figure~\ref{mixed_cases}) is shown to roughly approximate the effects not taken into account by the simple scaling, with the largest deviation to $18.3\%$. 

Equation (\ref{DQO_law}) is observed to have increasing accuracy as the Alfv\'en radii become larger in Figure~\ref{mixed_cases}, this is due to the increasing dominance of the dipolar component at large distances. Quantifying the scatter in our residual, we approximate the distribution of deviations as Gaussian, and calculate a standard deviation of $5.1\%$, when evaluating all 160 of our simulated cases. Considering the 32 three component cases, the standard deviation remains of the same order $5.2\%$, indicating the formulation maintains precision with the inclusion of all three antisymmetric components. The largest deviations from the predicted values belong to the dipole-octupole simulations, and these are observed within Figures \ref{Upsilon_DO} and \ref{Up_dip_DO}. In both Figures, and the residual, the predicted values are shown to under estimate the simulated values, for small average Alfv\'en radii, but with increasing field strength begin to over predict. The trends in the residual represents physics not incorporated into our approximate formula, and can be explained. The underestimation at first, is due to the sharpness of regime transition from the broken power law representation, in reality there is a smoother transition which is always larger than the break in power laws. This significantly impacts the dipole-octupole simulations as they most often probe this regime, as can be seen within Figure~\ref{Up_dip_DO}. For the dipole-octupole combinations, we propose this transition must be much more broad to match the deviations in the residual of Figure~\ref{mixed_cases}.

Equation (\ref{DQO_law}) represents an approximation to the impact of mixed geometry fields on the prediction of the average Alfv\'en radius. Our mixed cases are found to be well behaved and can all be predicted by this formulation within $\approx\pm20\%$ accuracy for the most deviant, the majority lie within $\approx\pm5\%$ accuracy. 

\subsection{Analysis of Previous Mixed Fields}

\begin{table*}
\caption{Comparison of results, $R_{\text{A}}/R_*|_{sim}$, from cases of \cite{reville2015effect} to the prediction of equation (\ref{DQO_law}), $R_{\text{A}}/R_*|_{FM18}$. }
\label{reville_pred}
\center
    \begin{tabular}{c|cccc}
        \hline\hline
        Object &  $\mathcal{R}_{\text{dip}}|\mathcal{R}_{\text{quad}}|\mathcal{R}_{\text{oct}}$ & $\Upsilon$ & $R_{\text{A}}/R_*|_{sim}$ & $R_{\text{A}}/R_*|_{FM18}$  \\ \hline
        Sun Min   &  $-0.47|0.03|-0.50$  & $812$ & $6.7$ & $6.74$             \\ 
        Sun Max       &  $0.13|0.73|0.14$  & $130$ & $3.3$ & $3.36$               \\   
        TYC-0486     &  $-0.10|0.79|-0.11$  & $17600$ & $7.7$ & $7.10$              \\   
        \hline
    \end{tabular}
\end{table*}

\cite{reville2015effect} presented mixed field simulations containing axisymmetric dipole, quadrupole and octupole components, based on observations of the Sun, at maximum and minimum of magnetic activity, along with a solar-like star TYC-0486. To further test our formulation, we use input parameters and results from Table 3 of \cite{reville2015effect} and predict values for the average Alfv\'en radii of the mixed cases produced in their work. We use equation (\ref{DQO_law}) with the fit constants from their lower temperature wind ($c_{\text{s}}/v_{\text{esc}}=0.222$) and manipulate the given field strengths into suitable $\mathcal{R}_l$ values. Results can be found in Table \ref{reville_pred}, and are shown in Figure~\ref{mixed_cases} with red squares. The predicted values for the Alfv\'en radii agree to better than $10\%$ precision. The largest deviation, $\approx8\%$, is for TYC-0486, which we accredit to the location of the predicted Alfv\'en radius falling in between regimes, at the break in the power law (almost governed by the dipole component only), where the broken power law approximation has the biggest error. 

Recent work by \cite{reville2017global}, presented 13 thermally driven wind simulations, in 3D, for the solar wind, using Wilcox Solar Observatory magnetograms, spanning the years 1989-2001. These simulations use the spherical harmonic coefficients derived from the magnetograms, up to $l=15$, including the non-axisymmetric modes. We predict the values of the average Alfv\'en radii using equation (\ref{DQO_law}), allowing the strength of any non-axisymmetric component to be added in quadrature with the axisymmetric component to produce representative strengths for the dipole, quadrupole and octupole components. For example, the dipole field strength is computed, 
\begin{equation}
B_*^{l=1}=\sqrt{(B^{l=1}_{m=-1})^2+(B^{l=1}_{m=0})^2+(B^{l=1}_{m=1})^2}.
\end{equation}
We obtained the field strengths for the dipole, quadrupole and octupole components of the magnetograms used in the simulations of \cite{reville2017global}, ignoring the higher order field componets (R\'eville, private communication 2017). The results from this are shown in Figure~\ref{mixed_cases} with magenta squares, and show a good agreement in most cases to the simulated values. However, we note that the Alfv\'en radii tabulated within \cite{reville2017global} are geometrically-averaged rather than torque-averaged, as used in this work (both scale with wind magnetisation in a similar manner). These values thus represent the average spherical radius for the Alfv\'en surface in their 3D simulations. The base wind temperature for their simulations is also cooler ($c_{\text{s}}/v_{\text{esc}}\approx0.234$) than in our simulations. Nevertheless, Figure~\ref{mixed_cases} shows good agreement with the predicted values, we calculate a standard deviation of $8.4\%$. If we apply an approximate correction to the spherical radii with a factor of 2/3 (due to the angular momentum lever arm being proportional to $r\sin\theta$) and use torque scaling coefficients fit to the lower temperature wind from \cite{pantolmos2017magnetic}, we find that all the magenta simulations fit within the $10\%$ precision, despite the inclusion of the non-axisymmetric components. This suggests equation (\ref{DQO_law}) can be used in cases with non-axisymmetric geometries in combination, but further study is required to test more fully.

\section{Analysis Based on Open Flux}
\subsection{Magnetic Flux Profiles}

   \begin{figure*}
   \centering
    \includegraphics[width=0.8\textwidth]{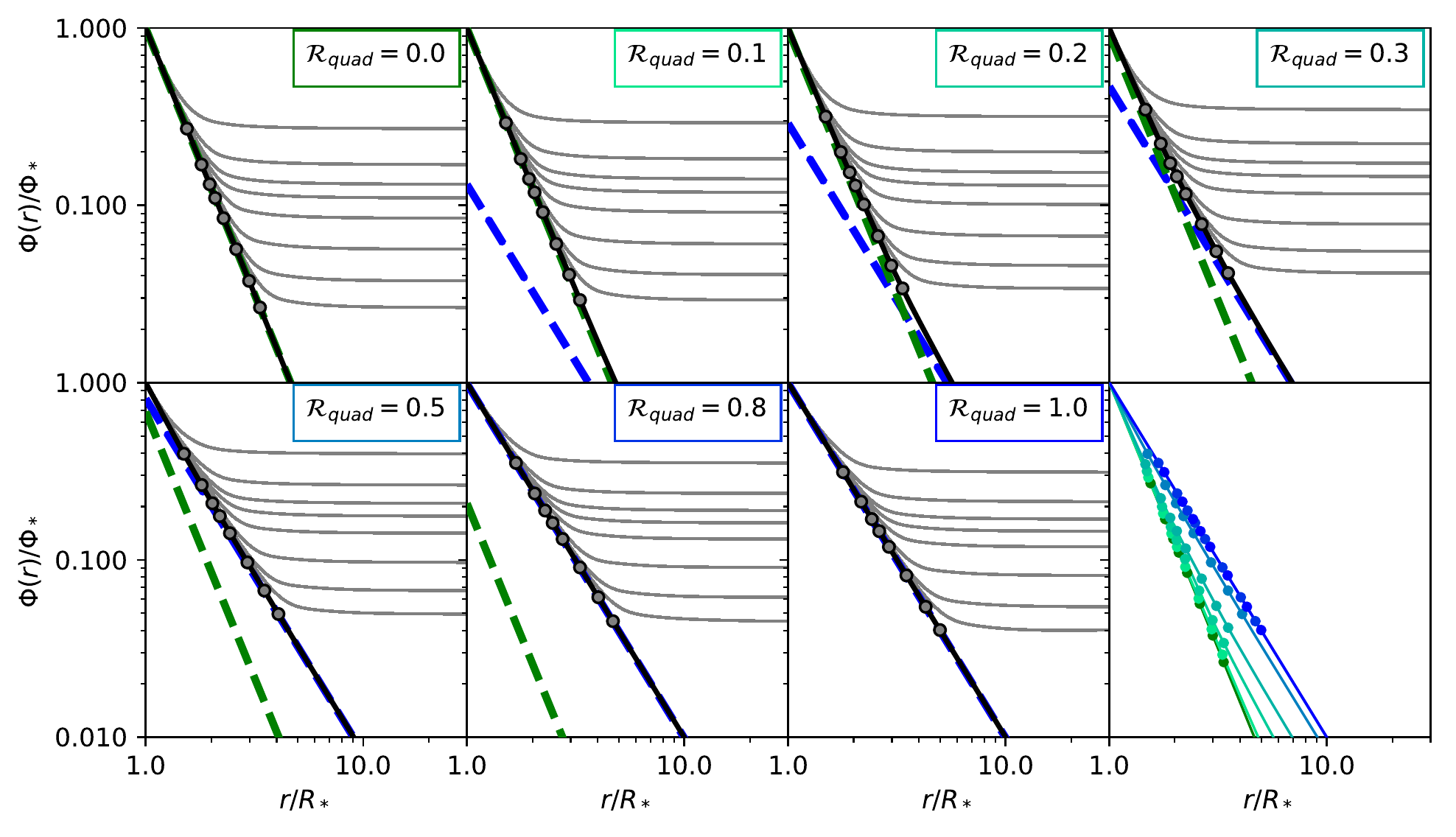}
     \caption{Unsigned magnetic flux vs radial distance (grey lines) for all the cases with combined quadrupole and octupole components (labelled $\mathcal{R}_{\text{quad}}=0.1-0.8$, along with the pure quadrupole and octupole cases (labelled $\mathcal{R}_{\text{quad}}=0.0$ and $1.0$). Thick dashed blue and green lines show the value for a potential field for the quadrupole and octupole component on their own. The total potential field flux, used as the initial condition, equation (\ref{phi}), is shown in solid black. Thin grey lines in each panel show the magnetic flux in a single steady state solution, for different field strengths of a given geometry. The flux within the simulations follows the potential field solution closely until the magnetic field is opened into a radial configuration with constant flux. Grey points indicate the location of the field opening radii $R_{\text{o}}$, as we define it in this work. The mixed field geometries decay with an octupolar dependence until reaching the quadrupolar component, at which point the quadrupole controls the decay. This explains why the broken power law approximation is a good fit to the data in most cases.  For comparison, the final panel shows all of the potential (initial) field geometries and their opening radii, coloured according to their $\mathcal{R}_{\text{quad}}$ value.}
     \label{QO_Flux}
  \end{figure*}
  
     \begin{figure*}
    \includegraphics[width=\textwidth]{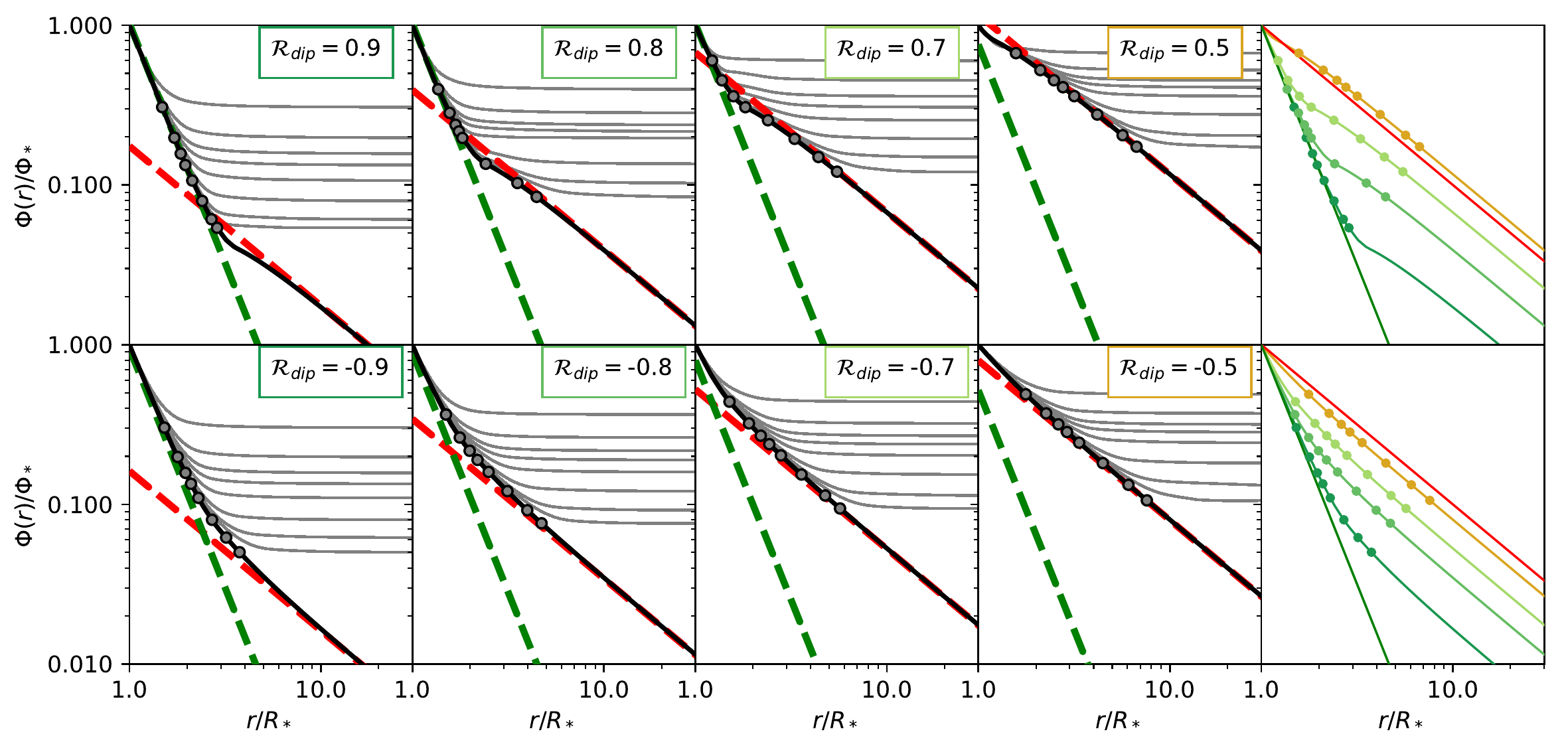}
     \caption{Unsigned magnetic flux vs radial distance for all the cases with combined dipole and octupole components (labelled $\mathcal{R}_{\text{dip}}=\pm0.5-0.9$), both aligned (top row) and anti-aligned (bottom row), in a similar format as figure \ref{QO_Flux}. Thick dashed red and green lines show the value for a potential field for the dipole and octupole component on their own. The aligned cases have a qualitatively different behaviour to the dipole-quadrupole, quadrupole-octupole and anti-aligned dipole-octupole cases, in that the former show a subtle inflection in the their flux vs radius (most apparent in the solid black lines for large $\mathcal{R}_{\text{dip}}$ values, the three top left panels). This is caused by the subtraction of the two fields in the equatorial region, which has a maximum effect at the radius where the two components have the same magnitude. The net effect of this inflection in the magnetic flux is subtle, and thus our scaling relation (which does not treat the aligned and anti-aligned cases differently) remains an acceptable approximation to all simulations. For comparison, the rightmost panel shows all of the potential (initial) field geometries and their opening radii, coloured according to their $\mathcal{R}_{\text{dip}}$ value, for the aligned and anti-aligned cases respectively.}
     \label{DO_Flux}
  \end{figure*}
  
The behaviour of the stellar wind torque, quantified in the previous sections, is similar to the results found in Paper I. Lower order magnetic components decay more slowly with radius than higher order components. Thus, the lower order component typically dominates the dynamics of the global torque. The higher order component can usually be ignored, unless it has a comparable field strength to the lower order component at the Alfv\'en radius, which requires the higher order field to dominate at the surface. 

The radial dependence of the magnetic field is best described by the unsigned magnetic flux. To calculate this, we evaluate an integral of the magnetic field threading closed spherical shells with area $A$, this produces the unsigned magnetic flux as a function of radial distance, 
\begin{equation}
\Phi(r)=\oint_r|{\bf B} \cdot d{\bf A}|.
\label{phi}
\end{equation}
For a potential field, as used in the initial conditions, the magnetic flux decays as a simple power law,
\begin{equation}
\Phi(r)=\Phi_*\bigg(\frac{R_*}{r}\bigg)^l
\end{equation} 
where $\Phi_*$ is the surface magnetic flux and $l$ represents the magnetic order of the field, increasing for more complex fields. Thus higher order fields decay radially faster.

\begin{figure*}
   \centering
    \includegraphics[width=\textwidth]{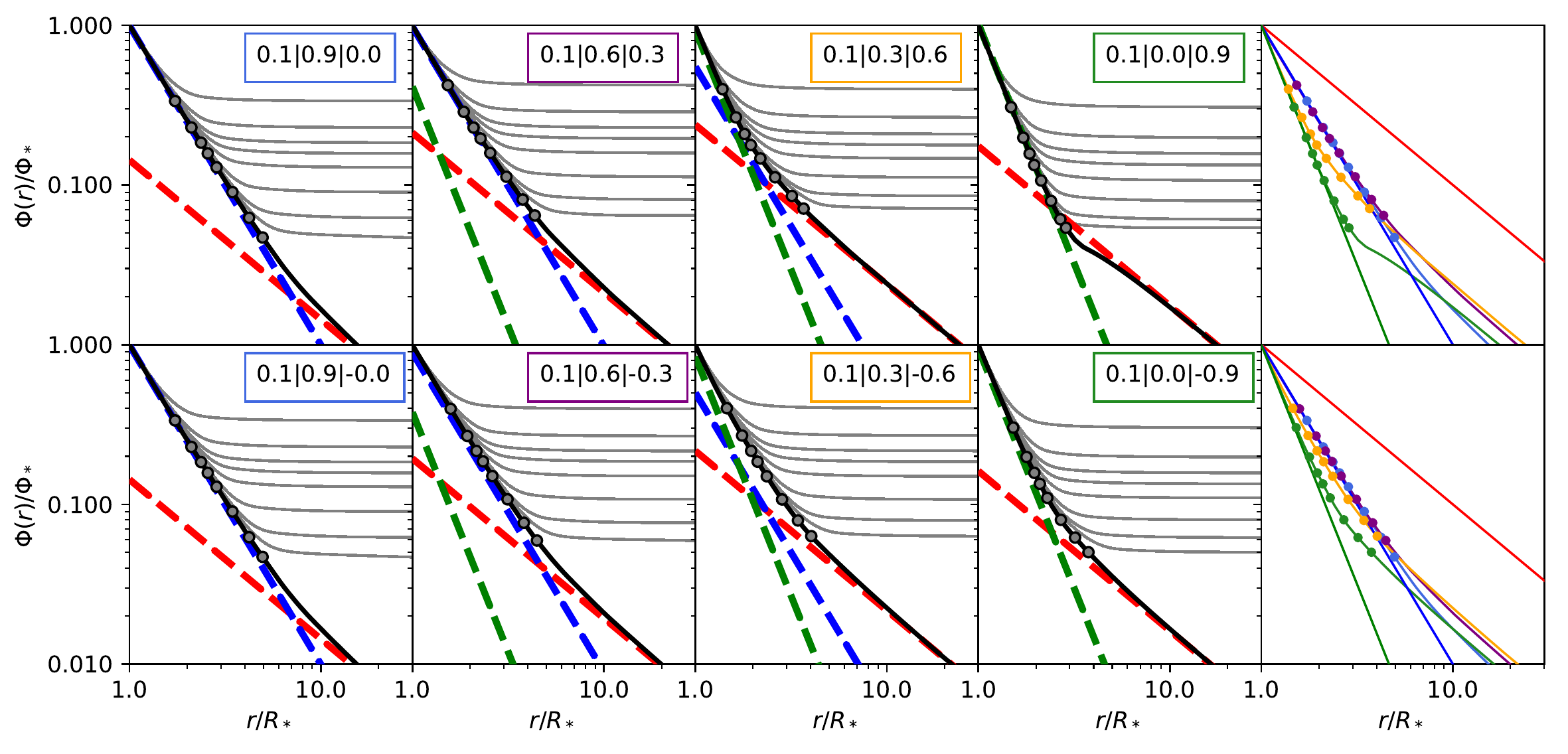}
     \caption{Unsigned magnetic flux vs radial distance for the sample of mixed dipole, quadrupole and octupole cases in the same format a Figure~\ref{QO_Flux}. All cases shown have $10\%$ in the dipole component. Then, from left to right, the fraction in the octupole increases from $0-90\%$ (with the remaining fraction in the quadrupole component). Top row has aligned dipole-octupole, the bottom has anti-aligned. }
     \label{Mixed_Flux}
  \end{figure*}

The radial profiles of the flux in our steady state solutions are shown with thin grey lines in Figures \ref{QO_Flux}, \ref{DO_Flux} and \ref{Mixed_Flux}. Each ratio ($\mathcal{R}_{l}$) represent a different combined field geometry with each grey line having a different field strength. In each figure we include the potential field solution for the flux with a solid black line, produced by equation (\ref{phi}), showing the initial magnetic field configuration. No longer is a single power law produced; instead the components interact and produce a varying radial decay. In magnetised winds, the magnetic forces balance the thermal and ram pressures close to the star. Therefore the unsigned flux approximately follows the potential solution. Further from the star the pressure of the wind forces the magnetic field into a nearly radial configuration, beyond which, the unsigned flux becomes constant. This constant value is referred to as the open flux, $\Phi_{\text{open}}$ (typically larger field strength produce a smaller fraction of open flux to surface flux).

In the cases with quadrupole-octupole mixed fields (Figure~\ref{QO_Flux}), the individual potential field quadrupole and octupole components are indicated with thick dashed blue and green lines respectively. As with the previous dipole and quadrupole addition, the broken power law behaviour shown in the Alfv\'en radius formulation is visible. The quadrupole component often represents the most significant contribution to the total flux, as the dipole did within Paper I. The lower right panel of Figure~\ref{QO_Flux} shows the relative decay of all the potential fields.
  
Figure~\ref{DO_Flux} shows the radial magnetic flux evolution for the dipole-octupole combinations in a similar format as Figure~\ref{QO_Flux}. A quantitatively similar behaviour to the dipole-quadrupole and quadrupole-octupole combinations is shown with the anti-aligned field geometries, seen in the bottom row. This explains why previously the anti-aligned cases provided a better fit to the broken power law approximation than the aligned cases. For the cases with an aligned octupole component, the profile of the flux decay is distinctly different. The smooth transition between the two regimes of the broken power law is replaced with a deviation from the dipole which passes below the dipole component at first, and then asymptotes back. This is caused by the subtraction of the dipole and octupole fields over the equator, which reduces the unsigned flux and has the largest impact at the radial distance where the two components have the equal and opposite field strength.

For these two component simulations, the approximate formulation, equation (\ref{DQO_law}), mathematically approximates the radial decay of the magnetic field with two regimes, an octupolar decay close in to the star followed by a sharp transition to the lower order geometry (dipole or quadrupole), which ignores any influence of the octupolar field. The formulation works well when this is a good approximation, which is typically the case for the dipole-quadrupole, quadrupole-octupole and anti-aligned dipole-octupole cases. The inflection of the magnetic flux for aligned cases creates a discrepancy between our simplification and the physics in the simulation, therefore we observe a scatter in our results between the aligned and anti-aligned cases. Our formulation is least precise when the inflection occurs near the Alfv\'en radius, causing the formula to over predict the average Alfv\'en radius. However, in Section 3.2.4 we show this to be a systematic and measurable effect, that does not impact the validity of equation (\ref{DQO_law}).

For the three component simulations, the behaviour of the dipole-octupolar component alignment is shown to oppose the previous dipole-octupole addition. Equation (\ref{DQO_law}) more accurately approximates the mixed field cases with an aligned octupole component, than with an anti-aligned component. To explore this we show the radial evolution of the magnetic flux in Figure~\ref{Mixed_Flux}. The top panel displays the aligned cases with increasing octupolar component and decreasing quadrupolar component, moving to the right. The reduction of flux, or inflection in the flux profile, due to the dipole and octupole addition is only seen to be significant for one case, where the octupole fraction is maximised. In the remaining cases the octupolar fraction is too small to produce a strong reduction in the equatorial flux with the dipole. Hence the well behaved relation between the simulated aligned cases and the predicted average Alfv\'en radii in Figure~\ref{mixed_cases}. The poorest fitting cases to equation (\ref{DQO_law}) are the anti-aligned mixed cases shown in Figure~\ref{Mixed_Flux} with purple and orange stars. The potential field solutions, shown with solid black lines, sit above the dashed component slopes (most significant for cases 153-160, in orange) showing an increased field strength due to the complex addition of the three components in combination. This is unlike most of the previous combined field cases, which are typically described by either one component or the other, hence the predicted values differ for these cases. 

This behaviour is difficult to parametrise within our Alfv\'en radius approximation as it requires knowledge about the magnetic field evolution in the wind. For this work, we simply show why the simulations deviate from equation (\ref{DQO_law}) and suggest care is taken when using such formulations with dipolar and octupolar components. 

\subsection{Open Flux Torque Relation}

\begin{table}
\caption{Open Flux Fit Parameters to equations (\ref{UP_OPEN_OLD}) \& (\ref{UP_OPEN})}
\label{fitValues_open}
\center
    \begin{tabular}{ccc}
        \hline\hline
        Topology($l)$ &  $K_{\text{o}}$& $m_{\text{o}}$   \\ \hline
        Dipole ($1$)    &  $0.33\pm0.03$  & $0.371\pm0.003$             \\ 
        Quadrupole ($2$)                & $0.63\pm0.02$  &$0.281\pm0.003$              \\   
        Octupole ($3$)                & $0.85\pm0.03$  &$0.227\pm0.004$              \\   
        All Simulations               & $0.46\pm0.03$  &$0.329\pm0.004$              \\   
        \hline
         &    $K_{\text{c}}$ & $m_{\text{c}}$   \\ \hline
        Topology Independent & $0.08\pm0.04$&$0.470\pm0.004$\\
        \hline
    \end{tabular}
\end{table}

   \begin{figure*}
    \includegraphics[width=\textwidth]{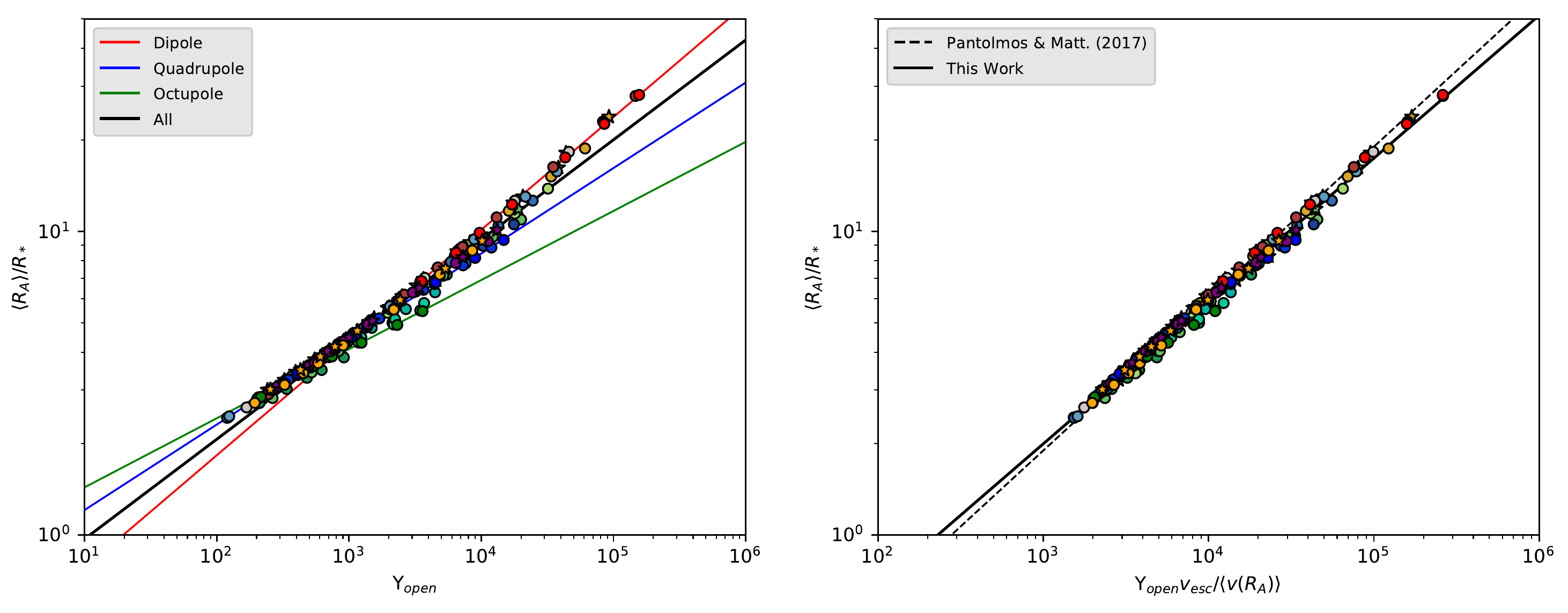}
     \caption{Average Alfv\'en radius vs the open flux magnetisation, $\Upsilon_{\text{open}}$, equation (\ref{upsilon_open}). All simulations from this study and Paper I are shown, colour-coded as in the previous figures. Left: Different scaling relations (equation \ref{UP_OPEN_OLD}, table \ref{fitValues_open}) are shown for each pure geometry and a combined fit. Right: Open Flux magnetisation divided by the average speed at the Alfv\'en surface $\langle v(R_{\text{A}}) \rangle$. The scatter is reduced, indicating that the different scalings in the left panel is primarily due to the effect of magnetic geometry on the wind acceleration (as discussed in Paper I). However there remains a small systematic trend, in that the higher order geometry winds sit lower for a given magnetisation (seen in Paper I), which may be due to systematic numerical effects. The solid black line represents the fit to all data (see table \ref{fitValues_open}), the dashed line represents the result from dipole wind simulations with different base wind temperatures from \cite{pantolmos2017magnetic}.} 
     \label{Up_Open_ALL}
  \end{figure*}

The open flux, $\Phi_{\text{open}}$, remains a key parameter in describing the torque scaling for any magnetic geometry. \cite{reville2015effect} constructs a semi-analytic formulation for the average Alfv\'en radius using the open flux wind magnetisation,
\begin{equation}
\Upsilon_{\text{open}}=\frac{\Phi_{\text{open}}^2/R_*^2}{\dot M v_{\text{esc}}}.
\label{upsilon_open}
\end{equation}
The dependence of the average Alfv\'en radius on $\Upsilon_{\text{open}}$ is then parametrised,
\begin{equation}
\frac{\langle R_{\text{A}}\rangle}{R_*} = K_{\text{o}}[\Upsilon_{\text{open}}]^{m_{\text{o}}},
\label{UP_OPEN_OLD}
\end{equation}
where $K_{\text{o}}$ and $m_{\text{o}}$ represent fit parameters to our simulations using this open flux formulation. In Paper I, we show the dependence of these fit parameters on magnetic geometry. We show this again within the left panel of Figure~\ref{Up_Open_ALL}. The scatter in average Alfv\'en radii values for different field geometries is reduced compared with that seen in the $\Upsilon$ parameter spaces (Figures \ref{Upsilon_DQ}, \ref{Upsilon_QO} and \ref{Upsilon_DO}), such that a single power law fit is viable, shown with a solid black line. However, better fits are obtained when considering each pure geometries independently, tabulated in Table \ref{fitValues_open}. 

Work by \cite{pantolmos2017magnetic} showed how differing wind acceleration affects the scaling relation by using different base wind temperatures to accelerate their winds. Different magnetic topologies produce slightly different wind acceleration from the stellar surface out to the Alfv\'en radius, due to the varying degree of super-radial expansion of the magnetic field lines \citep{velli2010solar, riley2012interplanetary, reville2016age}. Thus this causes the distinctly different scaling relations in the left panel of Figure~\ref{Up_Open_ALL}. Using the averaged Alfv\'en speed $\langle v(R_{\text{A}}) \rangle$ at the Alfv\'en surface, this difference in wind acceleration can be removed (see \citealp{pantolmos2017magnetic}), and the result is shown in the right panel of Figure~\ref{Up_Open_ALL}. 

The semi-analytic solution from \cite{pantolmos2017magnetic} is given by,
\begin{equation}
\frac{\langle R_{\text{A}}\rangle}{R_*} = K_{\text{c}}\bigg[\Upsilon_{\text{open}}\frac{ v_{\text{esc}}}{\langle v(R_{\text{A}})\rangle}\bigg]^{m_{\text{c}}},
\label{UP_OPEN}
\end{equation}
where $K_{\text{c}}$ and $m_{\text{c}}$ are fit parameters to this formulation. The fit relationship from \cite{pantolmos2017magnetic} and a fit to our simulation data (Table \ref{fitValues_open}), are shown with all our simulated cases (both Paper I \& II) in the right panel of Figure~\ref{Up_Open_ALL}. 

A small geometry dependent scatter remains in the right panel, which is noted in Paper I. The cause of which is an unanswered question, but may relate to systematic numerical errors due to modelling small scale complex field geometries. Our fit agrees well with that from \cite{pantolmos2017magnetic}, with a shallower slope due to the inclusion of the higher order geometries which show this systematic deviation from the dipole simulations. 

\subsection{Field Opening Radius}

     \begin{figure}
    \includegraphics[width=0.5\textwidth]{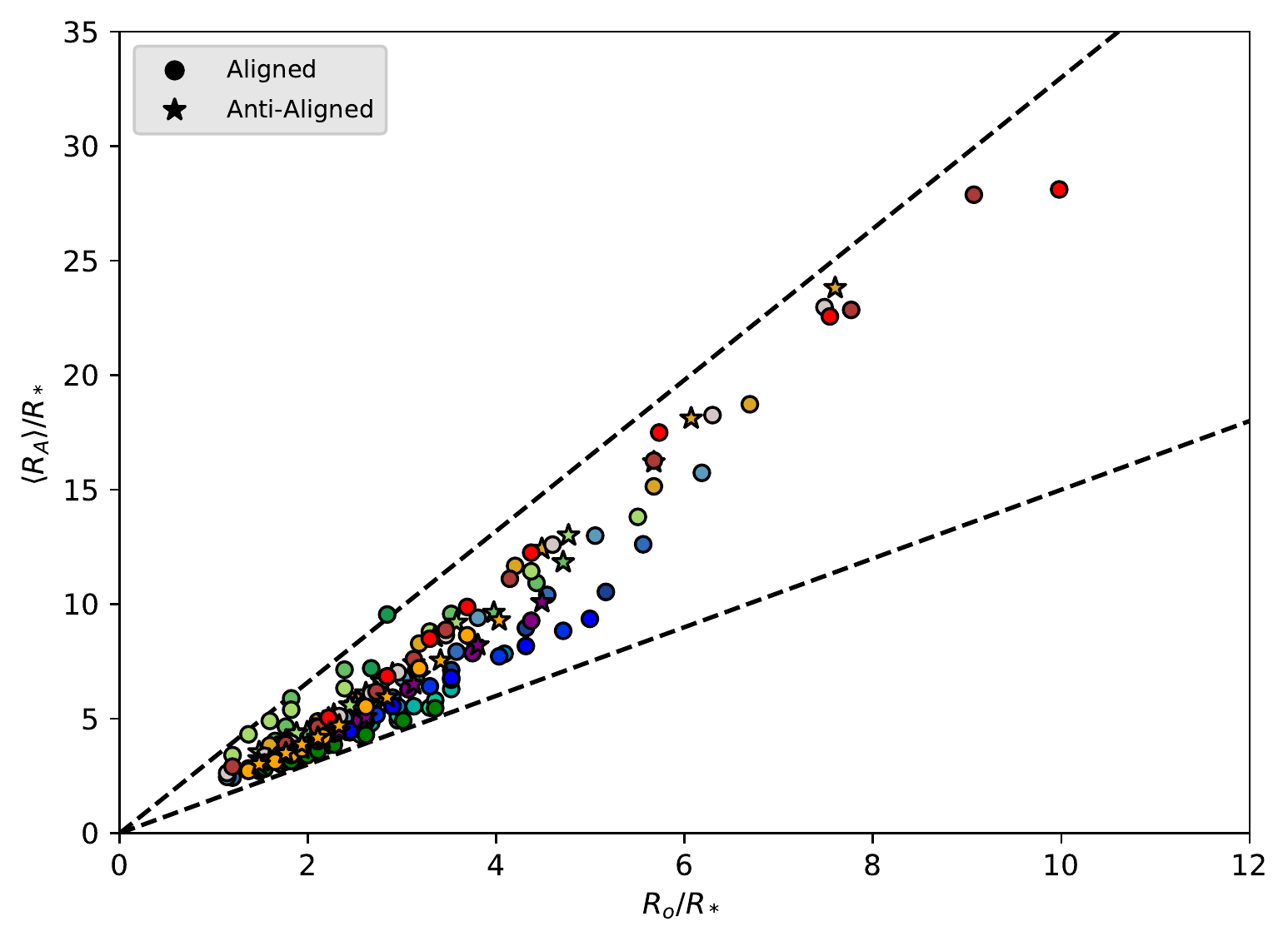}
     \caption{Average Alfv\'en radius vs opening radius for all cases. Black dashed lines represent $R_{\text{A}}/R_{\text{o}}=3.3$ and $1.5$, which bound all cases. The simulations show a similar behaviour to that discussed in Paper I, namely a geometry dependent separation, with the octupole geometries having the shallowest slope. }
     \label{RAvsRO}
  \end{figure}

As in previous works (e.g. \citealp{pantolmos2017magnetic}; Paper I), we define an opening radius $R_{\text{o}}$ using the value of the open flux. The opening radius is defined as the radial distance at which the potential field for a given geometry matches the value of the open flux, i.e. $\Phi(R_{\text{o}})=\Phi_{\text{open}}$. In this way, given the surface magnetic field geometry and the value of $R_{\text{o}}$, the open flux in the wind is recovered and thus the torque can be predicted. However, producing a single relation for predicting the opening radius, and thus the open flux, for our simulations remains an unsolved problem.

In Figures \ref{QO_Flux}, \ref{DO_Flux} and \ref{Mixed_Flux}, the opening radii for all simulations are marked with grey dots and compared in the final panel (coloured to match the respective $\mathcal{R}_l$ value). With increasing field strength, the simulations produce a larger average Alfv\'en radius and a larger deadzone/opening radius. The Alfv\'en and opening radii roughly grow together with increasing wind magnetisation, but their actual behaviour is more complex. The field complexity also has an affect on this relationship, with more complex geometries producing smaller opening radii, as the wind pressure is able to open the magnetic field closer to the star. 

We compare the average Alfv\'en radii and opening radii within Figure~\ref{RAvsRO}. The simulations containing an octupolar component, in general, show a shallower dependence, which continues the trend from dipole to quadrupole presented in Paper I. Interestingly, the aligned dipole-octuople fields are shown to have reduced values of $R_{\text{o}}$ for the Alfv\'en radii they produce, compared to the aligned cases, which is a consequence of the reduced flux from the field subtraction over the equator. For these cases the wind pressure iable to open the field much closer to the star, compared to the anti-aligned cases.

The relationship between the opening radius and the lever arm for magnetic braking torque in our wind simulations is evidently complex and interrelated with magnetic geometry, field strength and mass loss rate. The opening radius, as we define it here, is algebraically related to the source surface radius, $r_{\text{ss}}$, used within the Potential Field Source Surface (PFSS) models. As such the $R_{\text{o}}$ scales with $r_{\text{ss}}$ for a given field geometry, and its behaviour with increasing field strength should be accounted for within future PFSS models. 

\section{Conclusions}
This work present results from 160 new MHD simulations, and 50 previously discussed simulations from Paper I, which we use to disentangle the impacts of complex magnetic field geometries on the spin-down torque produced by magnetised stellar winds. Axisymmetric dipole, quadrupole and octupole fields are used to construct differing combined field geometries. We systematically vary the ratios, $\mathcal{R}_{\text{dip}}$, $\mathcal{R}_{\text{quad}}$ and $\mathcal{R}_{\text{oct}}$, of each field geometry with a range of total field strengths. Here we reinforce results from Paper I. With simple estimates using realistic magnetic field topologies (obtained from ZDI observations) and representative field strengths and mass loss rates for main sequence stars, the dipole component dominates the spin-down process, irrespective of the higher order components (Finley et al. in prep). The original formulation from \cite{matt2012magnetic} remains robust in most cases even for significantly non-dipolar fields. Combined with the work from \cite{pantolmos2017magnetic}, these formulations represent a strong foundation for predicting the stellar wind torques from a variety of cool stars with differing properties. 

We show the distinctly different changes to topology from our combined primary (dipole, octupole) and secondary (quadrupole) symmetry family fields. ``Primary'' being antisymmetric about the equator and ``secondary'' symmetric about the equator \citep{mcfadden1991reversals, derosa2012solar}. The addition of a primary and secondary fields produces an asymmetric field about the stellar equator, in contrast to the combination of two primary fields which maintains equatorial symmetry. However, the latter case breaks the degeneracy of the field alignment, producing two different topologies dependent on the relative orientation of the combined geometries. This is not the case for primary and secondary field addition, i.e. dipole-quadrupole and quadrupole-octupole, which produces the same global field reflected about the equator. 

The magnetic braking torque is shown, as in Paper I, to be largely dependent on the dominant lowest order field component. For observed field geometries this is, in general, dipolar in nature. We parametrise the torque from our mixed fields simulations based on the decay of the magnetic field. The average Alfv\'en radius, $\langle R_{\text{A}} \rangle$, is defined to represent a lever arm, or efficiency factor, for the torque, equation (\ref{torque}). From our simulated cases we produce an approximate formulation for the average Alfv\'en radius, equation (\ref{DQO_law}), where each $K_{\text{s}}$ and $m_{\text{s}}$ have tabulated values from our simulations in Table \ref{fitValues}. These values are temperature dependent, e.g. $\approx1.7$MK for a $1M_{\sun}$ star. In this formulation, the octupole geometry dominates the magnetic field close to the star, then decays radially leaving the quadrupole governing the radial decay of the field and finally the quadrupole decays leaving only the dipole component of the field. In each regime the strength of the field includes any component that is yet to decay away. 

Using this formula we are able to predict the torque in all of our simulations to $\approx20\%$ accuracy, with the majority predicted to with $\approx5\%$. This is then extended to mixed field simulations presented in \cite{reville2015effect} and \cite{reville2017global}. The formulation presented within this work remains an approximation, with a smoother transition from each regime observed with the simulations. This work represents a modification to existing torque formulations, which accounts for combined field geometries in a very general way. A key finding remains that the dipole component is able to account for the majority of the magnetic braking torque, in most cases. Thus previous works based on the assumption of the dipolar component being representative of the global field are validated. It is noted here however that it is the dipole component of the field not the total field strength which enters in the torque formulation, therefore it is import to decompose any observed field correctly to avoid miscalculation. 

In this study, as in the previous, we do not include the effects of rapid rotation or varying coronal temperatures. Prescriptions for rotational effects on the three pure geometries studied here are available \citep{matt2012magnetic, reville2015effect}, along with differing coronal temperatures for dipolar geometries \citep{pantolmos2017magnetic}. In general, differences in wind driving parameters and physics will introduce more deviation from equation (\ref{DQO_law}), however it is expected to remain valid.

Work remains in modelling the behaviour of non-axisymmetric components on the stellar wind environments surrounding sun-like and low mass stars, and the associated spin-down torques. Observed fields are shown to host a varied amount of non-axisymmetry \citep[e.g.][]{see2015energy}. Works including more complex coronal magnetic fields such as the inclusion of magnetic spots \citep[e.g.][]{cohen2009effect, garraffo2015dependence}, tilted magnetospheres \cite[e.g.][]{vidotto2010simulations} and using ZDI observations \citep[e.g.][]{vidotto2011understanding, vidotto2014stellar, alvarado2016simulating, nicholson2016temporal, garraffo2016space, reville2016age}, have shown the impact of specific cases but have yet to fully parametrise the variety of potential magnetic geometries. The relative orientation of some field combinations shown in this work have produced differences in the braking lever arm; therefore we expect the same to be true for non-axisymmetric geometries in combination. Since equation (\ref{DQO_law}) predicts the Alfv\'en radii from \cite{reville2017global} (Section 3.3), this suggest that our approximate formulation holds for non-axisymmetric components (using a quadrature addition of $\pm l$ components), but this remains to be validated.


%% Included in this acknowledgments section are examples of the
%% AASTeX hypertext markup commands. Use \url without the optional [HREF]
%% argument when you want to print the url directly in the text. Otherwise,
%% use either \url or \anchor, with the HREF as the first argument and the
%% text to be printed in the second.

\acknowledgments
Thanks for helpful discussions and technical advice from Georgios Pantolmos, Victor See, Victor R\'eville, Sasha Brun and Claudio Zanni.
This project has received funding from the European Research Council (ERC) under the European Union’s Horizon 2020 research and innovation programme (grant agreement No 682393).
We thank Andrea Mignone and others for the development and maintenance of the PLUTO code.
Figures within this work are produced using the python package matplotlib \citep{hunter2007matplotlib}.

\bibliographystyle{yahapj}
\bibliography{Paper2}

\begin{thebibliography}{}
\providecommand\natexlab[1]{#1}
\providecommand\JournalTitle[1]{#1}

\bibitem[{Ag{\"u}eros {et~al.}(2011)Ag{\"u}eros, Covey, Lemonias, Law, Kraus,
  Batalha, Bloom, Cenko, Kasliwal, Kulkarni, {et~al.}}]{agueros2011factory}
Ag{\"u}eros, M.~A., Covey, K.~R., Lemonias, J.~J., {et~al.} 2011,
  \JournalTitle{The Astrophysical Journal}, 740, 110

\bibitem[{Alvarado-G{\'o}mez {et~al.}(2016)Alvarado-G{\'o}mez, Hussain, Cohen,
  Drake, Garraffo, Grunhut, \& Gombosi}]{alvarado2016simulating}
Alvarado-G{\'o}mez, J., Hussain, G., Cohen, O., {et~al.} 2016,
  \JournalTitle{Astronomy \& Astrophysics}, 594, A95

\bibitem[{Amard {et~al.}(2016)Amard, Palacios, Charbonnel, Gallet, \&
  Bouvier}]{amard2016rotating}
Amard, L., Palacios, A., Charbonnel, C., Gallet, F., \& Bouvier, J. 2016,
  \JournalTitle{Astronomy \& Astrophysics}, 587, A105

\bibitem[{Aschwanden(2006)}]{aschwanden2006physics}
Aschwanden, M. 2006, Physics of the solar corona: an introduction with problems
  and solutions (Springer Science \& Business Media)

\bibitem[{Barnes(2003)}]{barnes2003rotational}
Barnes, S.~A. 2003, \JournalTitle{The Astrophysical Journal}, 586, 464

\bibitem[{Barnes(2010)}]{barnes2010simple}
---. 2010, \JournalTitle{The Astrophysical Journal}, 722, 222

\bibitem[{Blackman \& Owen(2016)}]{blackman2016minimalist}
Blackman, E.~G., \& Owen, J.~E. 2016, \JournalTitle{Monthly Notices of the
  Royal Astronomical Society}, 458, 1548

\bibitem[{Bouvier {et~al.}(2014)Bouvier, Matt, Mohanty, Scholz, Stassun, \&
  Zanni}]{bouvier2014angular}
Bouvier, J., Matt, S.~P., Mohanty, S., {et~al.} 2014, \JournalTitle{Protostars
  and Planets VI}, 433

\bibitem[{Brown(2014)}]{brown2014metastable}
Brown, T.~M. 2014, \JournalTitle{The Astrophysical Journal}, 789, 101

\bibitem[{Brun \& Browning(2017)}]{brun2017magnetism}
Brun, A.~S., \& Browning, M.~K. 2017, \JournalTitle{Living Reviews in Solar
  Physics}, 14, 4

\bibitem[{Brun {et~al.}(2013)Brun, Petit, Jardine, \&
  Spruit}]{brun2013rotation}
Brun, A.~S., Petit, P., Jardine, M., \& Spruit, H.~C. 2013,
  \JournalTitle{International Astronomical Union. Proceedings of the
  International Astronomical Union}, 9, 114

\bibitem[{Cohen {et~al.}(2009)Cohen, Drake, Kashyap, \&
  Gombosi}]{cohen2009effect}
Cohen, O., Drake, J., Kashyap, V., \& Gombosi, T. 2009, \JournalTitle{The
  Astrophysical Journal}, 699, 1501

\bibitem[{Cohen {et~al.}(2011)Cohen, Kashyap, Drake, Sokolov, Garraffo, \&
  Gombosi}]{cohen2011dynamics}
Cohen, O., Kashyap, V., Drake, J., {et~al.} 2011, \JournalTitle{The
  Astrophysical Journal}, 733, 67

\bibitem[{Cranmer \& Van~Ballegooijen(2005)}]{cranmer2005generation}
Cranmer, S., \& Van~Ballegooijen, A. 2005, \JournalTitle{The Astrophysical
  Journal Supplement Series}, 156, 265

\bibitem[{Cranmer {et~al.}(2017)Cranmer, Gibson, \& Riley}]{cranmer2017origins}
Cranmer, S.~R., Gibson, S.~E., \& Riley, P. 2017, \JournalTitle{Space Science
  Reviews}, 1

\bibitem[{Cranmer {et~al.}(2007)Cranmer, Van~Ballegooijen, \&
  Edgar}]{cranmer2007self}
Cranmer, S.~R., Van~Ballegooijen, A.~A., \& Edgar, R.~J. 2007,
  \JournalTitle{The Astrophysical Journal Supplement Series}, 171, 520

\bibitem[{{Davenport}(2017)}]{2017ApJ...835...16D}
{Davenport}, J.~R.~A. 2017,
  \href{http://dx.doi.org/10.3847/1538-4357/835/1/16}{\JournalTitle{\apj}, 835,
  16}

\bibitem[{Delorme {et~al.}(2011)Delorme, Cameron, Hebb, Rostron, Lister,
  Norton, Pollacco, \& West}]{delorme2011stellar}
Delorme, P., Cameron, A.~C., Hebb, L., {et~al.} 2011, \JournalTitle{Monthly
  Notices of the Royal Astronomical Society}, 413, 2218

\bibitem[{DeRosa {et~al.}(2012)DeRosa, Brun, \& Hoeksema}]{derosa2012solar}
DeRosa, M., Brun, A., \& Hoeksema, J. 2012, \JournalTitle{The Astrophysical
  Journal}, 757, 96

\bibitem[{do~Nascimento~Jr {et~al.}(2016)do~Nascimento~Jr, Vidotto, Petit,
  Folsom, Castro, Marsden, Morin, de~Mello, Meibom, Jeffers,
  {et~al.}}]{do2016magnetic}
do~Nascimento~Jr, J.-D., Vidotto, A., Petit, P., {et~al.} 2016,
  \JournalTitle{The Astrophysical Journal Letters}, 820, L15

\bibitem[{Donati {et~al.}(2006)Donati, Forveille, Cameron, Barnes, Delfosse,
  Jardine, \& Valenti}]{donati2006large}
Donati, J.-F., Forveille, T., Cameron, A.~C., {et~al.} 2006,
  \JournalTitle{Science}, 311, 633

\bibitem[{Donati {et~al.}(2007)Donati, Jardine, Gregory, Petit, Bouvier,
  Dougados, M{\'e}nard, Cameron, Harries, Jeffers,
  {et~al.}}]{donati2007magnetic}
Donati, J.-F., Jardine, M., Gregory, S., {et~al.} 2007, \JournalTitle{Monthly
  Notices of the Royal Astronomical Society}, 380, 1297

\bibitem[{Donati {et~al.}(2008)Donati, Moutou, Fares, Bohlender, Catala,
  Deleuil, Shkolnik, Cameron, Jardine, \& Walker}]{donati2008magnetic}
Donati, J.-F., Moutou, C., Fares, R., {et~al.} 2008, \JournalTitle{Monthly
  Notices of the Royal Astronomical Society}, 385, 1179

\bibitem[{Fares {et~al.}(2009)Fares, Donati, Moutou, Bohlender, Catala,
  Deleuil, Shkolnik, Cameron, Jardine, \& Walker}]{fares2009magnetic}
Fares, R., Donati, J.-F., Moutou, C., {et~al.} 2009, \JournalTitle{Monthly
  Notices of the Royal Astronomical Society}, 398, 1383

\bibitem[{Feldman {et~al.}(2005)Feldman, Landi, \&
  Schwadron}]{feldman2005sources}
Feldman, U., Landi, E., \& Schwadron, N. 2005, \JournalTitle{Journal of
  Geophysical Research: Space Physics}, 110

\bibitem[{Finley \& Matt(2017)}]{finley2017dipquad}
Finley, A.~J., \& Matt, S.~P. 2017,
  \href{http://stacks.iop.org/0004-637X/845/i=1/a=46}{\JournalTitle{The
  Astrophysical Journal}, 845, 46}

\bibitem[{Fisk {et~al.}(1998)Fisk, Schwadron, \& Zurbuchen}]{fisk1998slow}
Fisk, L., Schwadron, N., \& Zurbuchen, T. 1998, \JournalTitle{Space Science
  Reviews}, 86, 51

\bibitem[{Folsom {et~al.}(2016)Folsom, Petit, Bouvier, L{\`e}bre, Amard,
  Palacios, Morin, Donati, Jeffers, Marsden, {et~al.}}]{folsom2016evolution}
Folsom, C.~P., Petit, P., Bouvier, J., {et~al.} 2016, \JournalTitle{Monthly
  Notices of the Royal Astronomical Society}, 457, 580

\bibitem[{Gallet \& Bouvier(2013)}]{gallet2013improved}
Gallet, F., \& Bouvier, J. 2013, \JournalTitle{Astronomy \& Astrophysics}, 556,
  A36

\bibitem[{Gallet \& Bouvier(2015)}]{gallet2015improved}
---. 2015, \JournalTitle{Astronomy \& Astrophysics}, 577, A98

\bibitem[{Garraffo {et~al.}(2013)Garraffo, Cohen, Drake, \&
  Downs}]{garraffo2013effect}
Garraffo, C., Cohen, O., Drake, J., \& Downs, C. 2013, \JournalTitle{The
  Astrophysical Journal}, 764, 32

\bibitem[{Garraffo {et~al.}(2015)Garraffo, Drake, \&
  Cohen}]{garraffo2015dependence}
Garraffo, C., Drake, J.~J., \& Cohen, O. 2015, \JournalTitle{The Astrophysical
  Journal}, 813, 40

\bibitem[{Garraffo {et~al.}(2016{\natexlab{a}})Garraffo, Drake, \&
  Cohen}]{garraffo2016missing}
---. 2016{\natexlab{a}}, \JournalTitle{Astronomy \& Astrophysics}, 595, A110

\bibitem[{Garraffo {et~al.}(2016{\natexlab{b}})Garraffo, Drake, \&
  Cohen}]{garraffo2016space}
---. 2016{\natexlab{b}}, \JournalTitle{The Astrophysical Journal Letters}, 833,
  L4

\bibitem[{Gray(1984)}]{gray1984measurements}
Gray, D. 1984, \JournalTitle{The Astrophysical Journal}, 277, 640

\bibitem[{Gregory {et~al.}(2012)Gregory, Donati, Morin, Hussain, Mayne,
  Hillenbrand, \& Jardine}]{gregory2012can}
Gregory, S., Donati, J.-F., Morin, J., {et~al.} 2012, \JournalTitle{The
  Astrophysical Journal}, 755, 97

\bibitem[{Gregory \& Donati(2011)}]{gregory2011analytic}
Gregory, S.~G., \& Donati, J.-F. 2011, \JournalTitle{Astronomische
  Nachrichten}, 332, 1027

\bibitem[{Gregory {et~al.}(2016)Gregory, Donati, \&
  Hussain}]{gregory2016multipolar}
Gregory, S.~G., Donati, J.-F., \& Hussain, G.~A. 2016, \JournalTitle{arXiv
  preprint arXiv:1609.00273}

\bibitem[{G{\"u}del(2007)}]{gudel2007sun}
G{\"u}del, M. 2007, \JournalTitle{Living Reviews in Solar Physics}, 4, 3

\bibitem[{H{\'e}brard {et~al.}(2016)H{\'e}brard, Donati, Delfosse, Morin,
  Moutou, \& Boisse}]{hebrard2016modelling}
H{\'e}brard, {\'E}., Donati, J.-F., Delfosse, X., {et~al.} 2016,
  \JournalTitle{Monthly Notices of the Royal Astronomical Society}, 461, 1465

\bibitem[{Holzwarth(2005)}]{holzwarth2005impact}
Holzwarth, V. 2005, \JournalTitle{Astronomy \& Astrophysics}, 440, 411

\bibitem[{Hunter(2007)}]{hunter2007matplotlib}
Hunter, J.~D. 2007, \JournalTitle{Computing In Science \& Engineering}, 9, 90

\bibitem[{Hussain \& Alecian(2013)}]{hussain2013role}
Hussain, G.~A., \& Alecian, E. 2013, \JournalTitle{Proceedings of the
  International Astronomical Union}, 9, 25

\bibitem[{Hussain {et~al.}(2002)Hussain, Van~Ballegooijen, Jardine, \&
  Cameron}]{hussain2002coronal}
Hussain, G.~A., Van~Ballegooijen, A., Jardine, M., \& Cameron, A.~C. 2002,
  \JournalTitle{The Astrophysical Journal}, 575, 1078

\bibitem[{Irwin \& Bouvier(2009)}]{irwin2009ages}
Irwin, J., \& Bouvier, J. 2009, in IAU Symp, Vol. 258

\bibitem[{Jeffers {et~al.}(2014)Jeffers, Petit, Marsden, Morin, Donati, \&
  Folsom}]{jeffers2014e}
Jeffers, S., Petit, P., Marsden, S., {et~al.} 2014, \JournalTitle{Astronomy \&
  Astrophysics}, 569, A79

\bibitem[{Johns-Krull \& Valenti(2000)}]{johns2000measurements}
Johns-Krull, C.~M., \& Valenti, J.~A. 2000, in Stellar Clusters and
  Associations: Convection, Rotation, and Dynamos, Vol. 198, 371

\bibitem[{Kawaler(1988)}]{kawaler1988angular}
Kawaler, S.~D. 1988, \JournalTitle{The Astrophysical Journal}, 333, 236

\bibitem[{Keppens \& Goedbloed(1999)}]{keppens1999numerical}
Keppens, R., \& Goedbloed, J. 1999, \JournalTitle{Astron. Astrophys}, 343, 251

\bibitem[{Keppens \& Goedbloed(2000)}]{keppens2000stellar}
---. 2000, \JournalTitle{The Astrophysical Journal}, 530, 1036

\bibitem[{Kochukhov {et~al.}(2017)Kochukhov, Petit, Strassmeier, Carroll,
  Fares, Folsom, Jeffers, Korhonen, Monnier, Morin,
  {et~al.}}]{kochukhov2017surface}
Kochukhov, O., Petit, P., Strassmeier, K., {et~al.} 2017,
  \JournalTitle{Astronomische Nachrichten}, 338, 428

\bibitem[{Marcy(1984)}]{marcy1984observations}
Marcy, G. 1984, \JournalTitle{The Astrophysical Journal}, 276, 286

\bibitem[{Matt \& Pudritz(2008)}]{matt2008accretion}
Matt, S., \& Pudritz, R.~E. 2008, \JournalTitle{The Astrophysical Journal},
  678, 1109

\bibitem[{Matt {et~al.}(2015)Matt, Brun, Baraffe, Bouvier, \&
  Chabrier}]{matt2015mass}
Matt, S.~P., Brun, A.~S., Baraffe, I., Bouvier, J., \& Chabrier, G. 2015,
  \JournalTitle{The Astrophysical Journal Letters}, 799, L23

\bibitem[{Matt {et~al.}(2012)Matt, MacGregor, Pinsonneault, \&
  Greene}]{matt2012magnetic}
Matt, S.~P., MacGregor, K.~B., Pinsonneault, M.~H., \& Greene, T.~P. 2012,
  \JournalTitle{The Astrophysical Journal Letters}, 754, L26

\bibitem[{McFadden {et~al.}(1991)McFadden, Merrill, McElhinny, \&
  Lee}]{mcfadden1991reversals}
McFadden, P., Merrill, R., McElhinny, M., \& Lee, S. 1991,
  \JournalTitle{Journal of Geophysical Research: Solid Earth}, 96, 3923

\bibitem[{McQuillan {et~al.}(2013)McQuillan, Aigrain, \&
  Mazeh}]{mcquillan2013measuring}
McQuillan, A., Aigrain, S., \& Mazeh, T. 2013, \JournalTitle{Monthly Notices of
  the Royal Astronomical Society}, 432, 1203

\bibitem[{Meibom {et~al.}(2009)Meibom, Mathieu, \& Stassun}]{meibom2009stellar}
Meibom, S., Mathieu, R.~D., \& Stassun, K.~G. 2009, \JournalTitle{The
  Astrophysical Journal}, 695, 679

\bibitem[{Meibom {et~al.}(2011)Meibom, Mathieu, Stassun, Liebesny, \&
  Saar}]{meibom2011color}
Meibom, S., Mathieu, R.~D., Stassun, K.~G., Liebesny, P., \& Saar, S.~H. 2011,
  \JournalTitle{The Astrophysical Journal}, 733, 115

\bibitem[{Mestel(1968)}]{mestel1968magnetic}
Mestel, L. 1968, \JournalTitle{Monthly Notices of the Royal Astronomical
  Society}, 138, 359

\bibitem[{Mestel(1984)}]{mestel1984angular}
---. 1984, in Cool Stars, Stellar Systems, and the Sun (Springer), 49

\bibitem[{Mignone(2009)}]{mignone2009pluto}
Mignone, A. 2009, \JournalTitle{Memorie della Societa Astronomica Italiana
  Supplementi}, 13, 67

\bibitem[{Mignone {et~al.}(2007)Mignone, Bodo, Massaglia, Matsakos, Tesileanu,
  Zanni, \& Ferrari}]{mignone2007pluto}
Mignone, A., Bodo, G., Massaglia, S., {et~al.} 2007, \JournalTitle{The
  Astrophysical Journal Supplement Series}, 170, 228

\bibitem[{Morgenthaler {et~al.}(2011)Morgenthaler, Petit, Morin, Auri{\`e}re,
  Dintrans, Konstantinova-Antova, \& Marsden}]{morgenthaler2011direct}
Morgenthaler, A., Petit, P., Morin, J., {et~al.} 2011,
  \JournalTitle{Astronomische Nachrichten}, 332, 866

\bibitem[{Morin {et~al.}(2008{\natexlab{a}})Morin, Donati, Petit, Delfosse,
  Forveille, Albert, Auri{\`e}re, Cabanac, Dintrans, Fares,
  {et~al.}}]{morin2008large}
Morin, J., Donati, J.-F., Petit, P., {et~al.} 2008{\natexlab{a}},
  \JournalTitle{Monthly Notices of the Royal Astronomical Society}, 390, 567

\bibitem[{Morin {et~al.}(2008{\natexlab{b}})Morin, Donati, Forveille, Delfosse,
  Dobler, Petit, Jardine, Cameron, Albert, Manset, {et~al.}}]{morin2008stable}
Morin, J., Donati, J.-F., Forveille, T., {et~al.} 2008{\natexlab{b}},
  \JournalTitle{Monthly Notices of the Royal Astronomical Society}, 384, 77

\bibitem[{Nicholson {et~al.}(2016)Nicholson, Vidotto, Mengel, Brookshaw,
  Carter, Petit, Marsden, Jeffers, Fares, Collaboration,
  {et~al.}}]{nicholson2016temporal}
Nicholson, B., Vidotto, A., Mengel, M., {et~al.} 2016, \JournalTitle{Monthly
  Notices of the Royal Astronomical Society}, 459, 1907

\bibitem[{Pantolmos \& Matt(2017)}]{pantolmos2017magnetic}
Pantolmos, G., \& Matt, S.~P. 2017, \JournalTitle{The Astrophysical Journal},
  849, 83

\bibitem[{Parker(1965)}]{parker1965dynamical}
Parker, E. 1965, \JournalTitle{Space Science Reviews}, 4, 666

\bibitem[{Petit {et~al.}(2008)Petit, Dintrans, Solanki, Donati, Auri{\`e}re,
  Ligni{\`e}res, Morin, Paletou, Ramirez, Catala, {et~al.}}]{petit2008toroidal}
Petit, P., Dintrans, B., Solanki, S., {et~al.} 2008, \JournalTitle{Monthly
  Notices of the Royal Astronomical Society}, 388, 80

\bibitem[{Pinto {et~al.}(2016)Pinto, Brun, \& Rouillard}]{pinto2016flux}
Pinto, R., Brun, A., \& Rouillard, A. 2016, \JournalTitle{Astronomy \&
  Astrophysics}, 592, A65

\bibitem[{Pinto {et~al.}(2011)Pinto, Brun, Jouve, \&
  Grappin}]{pinto2011coupling}
Pinto, R.~F., Brun, A.~S., Jouve, L., \& Grappin, R. 2011, \JournalTitle{The
  Astrophysical Journal}, 737, 72

\bibitem[{Reiners(2012)}]{reiners2012observations}
Reiners, A. 2012, \JournalTitle{Living Reviews in Solar Physics}, 9, 1

\bibitem[{Reiners \& Mohanty(2012)}]{reiners2012radius}
Reiners, A., \& Mohanty, S. 2012, \JournalTitle{The Astrophysical Journal},
  746, 43

\bibitem[{R{\'e}ville \& Brun(2017)}]{reville2017global}
R{\'e}ville, V., \& Brun, A.~S. 2017, \JournalTitle{arXiv preprint
  arXiv:1710.02908}

\bibitem[{R{\'e}ville {et~al.}(2015)R{\'e}ville, Brun, Matt, Strugarek, \&
  Pinto}]{reville2015effect}
R{\'e}ville, V., Brun, A.~S., Matt, S.~P., Strugarek, A., \& Pinto, R.~F. 2015,
  \JournalTitle{The Astrophysical Journal}, 798, 116

\bibitem[{R{\'e}ville {et~al.}(2016)R{\'e}ville, Folsom, Strugarek, \&
  Brun}]{reville2016age}
R{\'e}ville, V., Folsom, C.~P., Strugarek, A., \& Brun, A.~S. 2016,
  \JournalTitle{The Astrophysical Journal}, 832, 145

\bibitem[{Riley {et~al.}(2006)Riley, Linker, Miki{\'c}, Lionello, Ledvina, \&
  Luhmann}]{riley2006comparison}
Riley, P., Linker, J., Miki{\'c}, Z., {et~al.} 2006, \JournalTitle{The
  Astrophysical Journal}, 653, 1510

\bibitem[{Riley \& Luhmann(2012)}]{riley2012interplanetary}
Riley, P., \& Luhmann, J. 2012, \JournalTitle{Solar Physics}, 277, 355

\bibitem[{Robinson {et~al.}(1980)Robinson, Worden, \&
  Harvey}]{robinson1980observations}
Robinson, R., Worden, S., \& Harvey, J. 1980, \JournalTitle{The Astrophysical
  Journal}, 236, L155

\bibitem[{Saar(1990)}]{saar1990magnetic}
Saar, S. 1990, in Symposium-International Astronomical Union, Vol. 138,
  Cambridge University Press, 427

\bibitem[{Saikia {et~al.}(2016)Saikia, Jeffers, Morin, Petit, Folsom, Marsden,
  Donati, Cameron, Hall, Perdelwitz, {et~al.}}]{saikia2016solar}
Saikia, S.~B., Jeffers, S., Morin, J., {et~al.} 2016, \JournalTitle{Astronomy
  \& Astrophysics}, 594, A29

\bibitem[{Schatzman(1962)}]{schatzman1962theory}
Schatzman, E. 1962, in Annales d'astrophysique, Vol.~25, 18

\bibitem[{See {et~al.}(2015)See, Jardine, Vidotto, Donati, Folsom, Saikia,
  Bouvier, Fares, Gregory, Hussain, {et~al.}}]{see2015energy}
See, V., Jardine, M., Vidotto, A., {et~al.} 2015, \JournalTitle{Monthly Notices
  of the Royal Astronomical Society}, 453, 4301

\bibitem[{See {et~al.}(2016)See, Jardine, Vidotto, Donati, Saikia, Bouvier,
  Fares, Folsom, Gregory, Hussain, {et~al.}}]{see2016connection}
---. 2016, \JournalTitle{Monthly Notices of the Royal Astronomical Society},
  462, 4442

\bibitem[{See {et~al.}(2017{\natexlab{a}})See, Jardine, Vidotto, Donati,
  Saikia, Fares, Folsom, Jeffers, Marsden, Morin, {et~al.}}]{see2017open}
---. 2017{\natexlab{a}}, \JournalTitle{Monthly Notices of the Royal
  Astronomical Society}

\bibitem[{See {et~al.}(2017{\natexlab{b}})See, Jardine, Vidotto, Donati,
  Saikia, Fares, Folsom, H{\'e}brard, Jeffers, Marsden,
  {et~al.}}]{see2016studying}
---. 2017{\natexlab{b}}, \JournalTitle{Monthly Notices of the Royal
  Astronomical Society}, stw3094

\bibitem[{Skumanich(1972)}]{skumanich1972time}
Skumanich, A. 1972, \JournalTitle{The Astrophysical Journal}, 171, 565

\bibitem[{Soderblom(1983)}]{soderblom1983rotational}
Soderblom, D. 1983, \JournalTitle{The Astrophysical Journal Supplement Series},
  53, 1

\bibitem[{Stauffer {et~al.}(2016)Stauffer, Rebull, Bouvier, Hillenbrand,
  Collier-Cameron, Pinsonneault, Aigrain, Barrado, Bouy, Ciardi,
  {et~al.}}]{stauffer2016rotation}
Stauffer, J., Rebull, L., Bouvier, J., {et~al.} 2016, \JournalTitle{The
  Astronomical Journal}, 152, 115

\bibitem[{Steinolfson \& Hundhausen(1988)}]{steinolfson1988density}
Steinolfson, R., \& Hundhausen, A. 1988, \JournalTitle{Journal of Geophysical
  Research: Space Physics}, 93, 14269

\bibitem[{Testa {et~al.}(2004)Testa, Drake, \& Peres}]{testa2004density}
Testa, P., Drake, J.~J., \& Peres, G. 2004, \JournalTitle{The Astrophysical
  Journal}, 617, 508

\bibitem[{Ud-Doula {et~al.}(2009)Ud-Doula, Owocki, \&
  Townsend}]{ud2009dynamical}
Ud-Doula, A., Owocki, S.~P., \& Townsend, R.~H. 2009, \JournalTitle{Monthly
  Notices of the Royal Astronomical Society}, 392, 1022

\bibitem[{van~der Holst {et~al.}(2014)van~der Holst, Sokolov, Meng, Jin,
  Manchester~IV, T{\'o}th, \& Gombosi}]{van2014alfven}
van~der Holst, B., Sokolov, I.~V., Meng, X., {et~al.} 2014, \JournalTitle{The
  Astrophysical Journal}, 782, 81

\bibitem[{Van~Saders \& Pinsonneault(2013)}]{van2013fast}
Van~Saders, J.~L., \& Pinsonneault, M.~H. 2013, \JournalTitle{The Astrophysical
  Journal}, 776, 67

\bibitem[{Velli(2010)}]{velli2010solar}
Velli, M. 2010, in AIP Conference Proceedings, Vol. 1216, AIP, 14

\bibitem[{Vidotto {et~al.}(2011)Vidotto, Jardine, Opher, Donati, \&
  Gombosi}]{vidotto2011understanding}
Vidotto, A., Jardine, M., Opher, M., Donati, J., \& Gombosi, T. 2011, in 16th
  Cambridge Workshop on Cool Stars, Stellar Systems, and the Sun, Vol. 448,
  1293

\bibitem[{Vidotto {et~al.}(2009)Vidotto, Opher, Jatenco-Pereira, \&
  Gombosi}]{vidotto2009three}
Vidotto, A., Opher, M., Jatenco-Pereira, V., \& Gombosi, T. 2009,
  \JournalTitle{The Astrophysical Journal}, 699, 441

\bibitem[{Vidotto {et~al.}(2010)Vidotto, Opher, Jatenco-Pereira, \&
  Gombosi}]{vidotto2010simulations}
---. 2010, \JournalTitle{The Astrophysical Journal}, 720, 1262

\bibitem[{Vidotto {et~al.}(2014)Vidotto, Gregory, Jardine, Donati, Petit,
  Morin, Folsom, Bouvier, Cameron, Hussain, {et~al.}}]{vidotto2014stellar}
Vidotto, A., Gregory, S., Jardine, M., {et~al.} 2014, \JournalTitle{Monthly
  Notices of the Royal Astronomical Society}, 441, 2361

\bibitem[{Weber \& Davis(1967)}]{weber1967angular}
Weber, E.~J., \& Davis, L. 1967, \JournalTitle{The Astrophysical Journal}, 148,
  217

\end{thebibliography}

%% Tables may also be prepared as separate files. See the accompanying
%% sample file table.tex for an example of an external table file.
%% To include an external file in your main document, use the \input
%% command. Uncomment the line below to include table.tex in this
%% sample file. (Note that you will need to comment out the \documentclass,
%% \begin{document}, and \end{document} commands from table.tex if you want
%% to include it in this document.)


%% The following command ends your manuscript. LaTeX will ignore any text

\end{document}